\documentclass{WileyMSP-template}
\usepackage{amsmath,amsfonts}
\usepackage{algorithmic}
\usepackage{algorithm}
\usepackage{array}
\usepackage{textcomp}
\usepackage{stfloats}
\usepackage{url}
\usepackage{verbatim}
\usepackage{graphicx}
\usepackage[superscript]{cite}
\usepackage{dcolumn}
\usepackage{bm}
\usepackage[utf8]{inputenc}
\usepackage[T1]{fontenc}
\usepackage{mathptmx}
\usepackage{etoolbox}
\usepackage{subcaption}
\usepackage{makecell}
\usepackage{booktabs}
\usepackage{multirow}
\usepackage{xcolor}

\renewenvironment{affiliations}{%
  \par\medskip\large\setlength{\parindent}{0pt}%
}{\par}

\begin{document}

\title{Overcoming Transport Layer Bottlenecks to Quantify Ionic \\ Parameters from Transient Ion Current Measurements of Perovskite Solar Cells}

\maketitle


\author{Shudi Jiao\textsuperscript{1,2,\dag},}
\author{Miguel Torre Cachafeiro\textsuperscript{2,\dag},}
\author{Huagui Lai\textsuperscript{2},}
\author{Fuxiang Ji\textsuperscript{2},}
\author{Tristan Sachsenweger Ballantyne\textsuperscript{2,3},}
\author{Sharun Parayil Shaji\textsuperscript{2,4},}
\author{Matthias Diethelm\textsuperscript{5},}
\author{Fan Fu\textsuperscript{6},}
\author{Wei E.I. Sha\textsuperscript{1,*},} and
\author{Wolfgang Tress\textsuperscript{2,7,*}}

\begin{affiliations}
\normalsize
\textsuperscript{1}College of Information Science and Electronic Engineering, Zhejiang University, Hangzhou 310058, China\\
\textsuperscript{2}Institute of Computational Physics, Zurich University of Applied Sciences (ZHAW), 8400 Winterthur, Switzerland\\
\textsuperscript{3}Department of Chemistry and Applied Biosciences, ETH Zürich, Zürich 8093, Switzerland\\
\textsuperscript{4}Department of Mathematical Modeling and Machine Learning, University of Zurich, Zurich, Switzerland\\
\textsuperscript{5}Fluxim AG, Katharina-Sulzer-Platz 2, 8400 Winterthur, Switzerland\\
\textsuperscript{6}Laboratory for Thin Films and Photovoltaics, Empa -- Swiss Federal Laboratories for Materials Science and Technology, Dübendorf 8600, Switzerland\\
\textsuperscript{7}Physik-Institut, University of Zurich, 8006 Zurich, Switzerland\\[4pt]
\textsuperscript{\dag}These authors contributed equally to this work.\\
\textsuperscript{*}Corresponding authors: Wei E.I. Sha (weisha@zju.edu.cn); Wolfgang Tress (wolfgang.tress@zhaw.ch)
\end{affiliations}

\justifying


\begin{abstract}

In perovskite solar cells (PSCs), voltage step-induced transient ion current (TIC) measurements, commonly referred to as bias-assisted charge extraction (BACE), are frequently used to quantify ion density. Drift-diffusion simulations predict that the ion density computed from TIC saturates once mobile ions screen the electric field in the perovskite. However, experimental studies often report ion densities orders of magnitude above this limit, whose physical origin remains incompletely explained in terms of transport layer (TL) properties. In this work, the capacitance of the TLs is identified to be the fundamental bottleneck: the maximum quantifiable ion density is limited to the charge that can accumulate at the perovskite/TL interfaces, so that TIC most often depends more strongly on TL properties than on the ionic properties of the perovskite. Experiments with systematically varied C$_{\rm 60}$ electron-TL thickness (p-i-n) and Spiro-OMeTAD hole-TL doping (n-i-p) confirm this dependence across architectures. To overcome this limitation, the importance of a correction based on the average ionic displacement is discussed, and it is shown how extrapolating the TL-thickness trend towards the TL-free situation yields the actual ionic conductivity of the absorber, alongside density and mobility, depending on the assumed ionic model. Simulations are also examined in which ion penetration into the TLs or initial accumulation under forward bias raise the capacitive limit and extend TIC sensitivity to higher ion densities. The slow release of trapped carriers is also discussed as a potential source of current which can inflate the TIC signal. Overall, the presented analysis provides important practical considerations for interpreting TIC and quantifying ionic properties in PSCs.

\end{abstract}


\keywords{Perovskite, Ion migration, Ionic conductivity, Ion density, Transient ion current, Bias-assisted charge extraction}

\section{Introduction}

Metal halide perovskite solar cells (PSCs) exhibit pronounced ionic motion under electric fields, which causes transient responses and screening effects influencing the current density-voltage ($J$-$V$) curve\cite{tress2015understanding} and long-term operational stability.\cite{thiesbrummel2024ion} 
The screening of the electric field in the perovskite by mobile ions can lower the collection efficiency of photogenerated charges, leading to a significant short-circuit current decay if diffusive transport is not efficient.\cite{cachafeiro2025visualising} The interplay between ion density and electronic properties can accelerate degradation\cite{torre2025ionic}. Moreover, secondary electrochemical processes at the interfaces and contacts can generate or release additional ionic species under operation, which could redistribute through the device and may even migrate into the transport layers (TLs).\cite{bitton2023perovskite,moia2026defect,bertoluzzi2021incorporating}

Accurate quantification of the mobile ion density ($N_\mathrm{ion}$) is therefore essential to understand PSC device physics, benchmark materials, and optimize interfaces.\cite{dunfield2020defects} Consequently, researchers have employed a variety of electrical characterization techniques to estimate $N_\mathrm{ion}$ in PSCs; e.g., capacitance-frequency spectra,\cite{reichert2020probing,schmidt2024consistent} Mott–Schottky analysis,\cite{diekmann2023determination,diethelm2025probing} capacitance transients,\cite{futscher2019quantification} open-circuit voltage decays\cite{fischer2021assigning} and current transients.\cite{le2022quantification,schmidt2025quantification,diethelm2025probing} However, all such electrical measurements face fundamental limitations arising from the coupling of ionic and electronic properties, as well as the electrostatic limitations imposed by the device architecture, in particular the TLs.\cite{diekmann2023determination,schmidt2025many,schmidt2025characterization}

Transient ion current (TIC) measurements, also more commonly referred to as bias-assisted charge extraction (BACE), have become one of the most widely used approaches for comparing ion density across devices and tracking its evolution during degradation. \cite{thiesbrummel2024ion,seid2025mitigating,er2025electron,sowmeeh2025minimizing,kalasariya2026halide} Here, we refer to this technique as TIC rather than BACE, since ionic charges are not actually extracted in this case.\cite{kniepert2019reliability} To carry out this technique, the device is first held at a positive preconditioning voltage $V_{\mathrm{pre}}$ in the dark with the goal of minimizing the electric field in the perovskite originating from the built-in potential from electrodes or doped TLs with different work functions. 

Ideally, $V_\mathrm{pre}$ is set to the perovskite field-free or flat-band condition\cite{hart2024more}, i.e. the applied voltage that
minimizes the net ionic space charge in the absorber (here referred to as $V_{\mathrm{\min},Q}$), therefore making positive and negative ions distribute rather homogeneously in the bulk of the perovskite.\cite{torre2025ionic} Afterwards, the voltage is rapidly switched, commonly to 0\,V, exposing the device to the built-in potential in the first moments. This makes the mobile ions respond and drift towards the TLs with the effect of screening the field in the perovskite. Thus the resulting external current density $J$ comprises an ionic ($J_\mathrm{ion}$) and a displacement ($J_\mathrm{disp}$) component, so that $J(t) = J_\mathrm{ion}(t) + J_\mathrm{disp}(t)$, which oppose each other due to the field screening (on the timescale of ion migration, $J_\mathrm{disp}$ is negative). Integrating the ionic current density transient over time yields the total ionic charge displaced per unit area, $Q = \int_0^\infty \left[J(t) - J_\mathrm{disp}(t)\right] dt$, from which the mobile ion density can be inferred as
\begin{equation}
    N_\mathrm{ion} = \frac{1}{q \cdot f_d \cdot d_\mathrm{PSK}}
    \int_0^\infty \left[ J(t) - J_\mathrm{disp}(t) \right] \mathrm{d}t,
    \label{eq:Nion}
\end{equation}
where $q$ is the elementary charge, $d_\mathrm{PSK}$ is the perovskite (PSK) layer thickness and $f_d$ is a dimensionless prefactor such that the mean displacement of the ionic distribution is $\Delta \bar{x} = f_d \cdot d_\mathrm{PSK}$ (typically $f_d = 0.5$ or $1$, corresponding to the overall ion density being shifted by half or the full perovskite thickness, respectively).\cite{diekmann2023determination,diethelm2025probing,thiesbrummel2024ion} If $J_\mathrm{ion} \gg J_\mathrm{disp}$, the displacement current density term can be ignored in Eq.\,\ref{eq:Nion}, as commonly done.

Despite its increasing adoption as a quantitative probe of mobile ionic species, the ion densities derived from TIC measurements span several orders of magnitude (as do the timescales considered), even for nominally similar perovskite compositions. Such substantial discrepancies raise critical concerns regarding the absolute accuracy and reliability of the technique, and suggest that its quantitative interpretation remains insufficiently understood. Simulation studies have argued that ionic quantification becomes inherently limited when the mobile ion density exceeds the electrode charge density\cite{diekmann2023determination} or when operating in the electric field-limited or high ion density regime \cite{schmidt2025quantification,diethelm2025probing}, which for typical device simulations with TLs often happens roughly around $N_{\mathrm{ion}}\approx10^{16}-10^{17}$\,cm$^{-3}$. Once the electric field is fully screened by the ions, the measured current does not directly reflect the available mobile ion population.\cite{diethelm2025probing} In principle, no electrical measurement can accurately quantify ion densities in the field-limited regime.\cite{schmidt2025many,schmidt2025quantification}

Conversely, experimental studies have reported ion densities exceeding $10^{18}$\,cm$^{-3}$ from the same technique,\cite{peterson2025reliable} with such values typically emerging after aging.\cite{thiesbrummel2024ion,tayagaki2024ion,schmidt2025quantification} However, according to simulations of complete device stacks, this is a change that would, in principle, be undetectable with most electrical measurements. This discrepancy points to a fundamental gap in understanding what physically limits the TIC. In particular, the role of the TLs, which strongly influence device electrostatics and thus the ionic response, has not been systematically studied.

In this work, we show how the ion density quantified by TIC measurements is fundamentally limited by the TL properties. Using drift-diffusion simulations, we confirm that the $N_{\mathrm{ion}}$ determined from TIC can be largely insensitive to the actual ionic concentration of the perovskite absorber. Instead, we show that the maximum quantifiable $N_\mathrm{ion}$ from TIC is limited by the capacitance of the TLs ($C_{\rm TL}$). While previous studies have attempted to correct $N_\mathrm{ion}$ by estimating the potential drop across the TLs,\cite{schmidt2025characterization} we focus on a correction factor $f_d$ to capture how many ions move based on how much the overall ion density shifts, which in turn depends on $C_{\rm TL}$ and $N_\mathrm{ion}$. In simulations, we calculate $f_d$ from the average ionic displacement, given by the shift of the density-weighted centroid ($\bar{x}$) of each distribution. As the ions redistribute to screen the field,
\begin{equation}
\begin{split}
    f_d
    = \frac{\Delta\bar{x}}{d_{\mathrm{PSK}}}
    &= \frac{1}{d_{\mathrm{PSK}}}\sum_{i=\mathrm{c,a}} s_i\!\left(
       \frac{\int_0^{d_{\mathrm{PSK}}} x\, n_i(x,t_{\mathrm{final}})\,dx}
            {\int_0^{d_{\mathrm{PSK}}} n_i(x,t_{\mathrm{final}})\,dx}
     - \frac{\int_0^{d_{\mathrm{PSK}}} x\, n_i(x,t_{\mathrm{initial}})\,dx}
            {\int_0^{d_{\mathrm{PSK}}} n_i(x,t_{\mathrm{initial}})\,dx}\right) \\[4pt]
    &\approx \frac{1}{N_{\mathrm{ion}}\, d_{\mathrm{PSK}}}
             \sum_{i=\mathrm{c,a}} s_i \int_0^{d_{\mathrm{PSK}}} x\, n_i(x,t_{\mathrm{final}})\,dx,
\end{split}
\label{eq:CoM}
\end{equation}
where $n_i(x,t)$ is the distribution of species $i$ with conserved density $N_{\mathrm{ion}}$, $s_i=\pm1$ its charge sign (cations $+$, anions $-$), and $x$ the position in the perovskite layer, bounded by $x=0$ and $x=d_{\mathrm{PSK}}$; $t_{\mathrm{initial}}$ is the state before ions respond, with the field still unscreened, and $t_{\mathrm{final}}$ the screened steady state. Since cations and anions drift in opposite directions, the sign weighting makes their displacements add constructively. In the simulation literature, mobile ionic species in PSK are typically modeled in one of two ways. In the one-ion model, mobile cations (iodide vacancies $V_\mathrm{I}^+$) are compensated by a static, uniform negative countercharge; this is well justified on the timescales relevant to common transient measurements.\cite{richardson2016can,courtier2019transport,bertoluzzi2020mobile,schmidt2025characterization} However, migration of additional anionic species has also been reported experimentally (albeit generally on slower timescales),\cite{zhao2017mobile,li2019cation,azpiroz2015defect} and atomistic simulations show iodide interstitials ($I_\mathrm{i}^-$) in CsPbI\textsubscript{3} with bulk mobilities comparable to $V_\mathrm{I}^+$ at room temperature.\cite{tyagi2025tracing} This motivates the two-ion model where both cations and anions redistribute, an approach also widely used for PSC simulations.\cite{van2015modeling,walter2018transient,messmer2025toward}

In the ideal case where all ions are preconditioned into a uniform bulk distribution at $t_{\mathrm{initial}}$, each species centroid starts at $d_{\mathrm{PSK}}/2$ and $f_d$ simplifies to the final-state ($t_{\mathrm{final}}$) expression in the second line of Eq.\,\eqref{eq:CoM}. $f_d$ naturally captures the main limiting behaviors: when all mobile ions drift toward the interfaces and the bulk becomes depleted, $f_d \to 1$ since both species move by $d_{\mathrm{PSK}}/2$ in opposite directions ($f_d \to 0.5$ if only one species is mobile); conversely, when the bulk remains compensated by equal densities of anions and cations and the net charge is confined to thin interfacial layers, the centroids barely shift and $f_d \to 0$. Here we show that the capacitance of the TLs limits ionic accumulation at the interfaces and thus the quantification of ionic parameters from TIC. Nevertheless, when used in Eq.\,\eqref{eq:Nion}, $f_d$ provides an accurate correction to obtain the true $N_{\mathrm{ion}}$; the need for this correction can be overcome by varying the TL thickness and extrapolating toward the TL-free limit. In the TL-free case, $f_d \to 1$ for two mobile ions independent of $N_\mathrm{ion}$. However, with one mobile ion only, compensated by a constant immobile background charge that prevents full depletion of the bulk, $f_d$ can fall below 0.5 (depending on the input $N_\mathrm{ion}$). An additional correction is therefore needed for the TL-free one-ion model, which can be calculated from the ionic depletion layer width within the perovskite.\cite{garcia2025ionic}

To validate the simulation results, we performed TIC measurements on lead-based PSCs in both p-i-n and n-i-p configurations, systematically varying the thickness of the C$_{60}$ electron transport layer (ETL) of p-i-n devices and the doping density of the Spiro-OMeTAD hole transport layer (HTL) of n-i-p devices. The experimental observations closely follow the simulation results in capturing the dependence on TL parameters. With the p-i-n devices, we perform an extrapolation of the trends in capacitance-dominated TIC towards an unlimited TL-free situation, which enables the quantification of the actual ionic conductivity, as well as the density and mobility (which vary depending on whether the one- or two-ion model is assumed).

For n-i-p devices with thick TLs, the displaced charge density and thus computed $N_{\rm ion}$ is found to be relatively high (close to $10^{18}$\,cm$^{-3}$ in some cases), exceeding typical simulated upper limits even without accounting for changes in $f_d$. To propose a plausible explanation for this discrepancy, we perform further simulations comprising a permeable model where ions are no longer confined in the perovskite but can penetrate the TLs to reach the electrodes and a model with traps at the perovskite/TL interfaces. Additionally, we discuss how preconditioning under high forward bias with an injection barrier can lead to substantial initial ionic accumulation that must first discharge during the transient. In all of these models, the computed $N_{\rm ion}$ can increase significantly above the capacitive limit, offering compelling explanations for the anomalously high $N_{\rm ion}$ values often reported from TIC measurements. The findings elucidate possible physical origins of underestimating ion density in TIC and other electrical measurements and offer key considerations for the accurate quantification of ionic parameters in PSC devices.

\section{Results and discussion}

\subsection{The bottlenecks in ionic quantification}

\begin{figure}[h]
\centering
  \includegraphics[width=\linewidth]{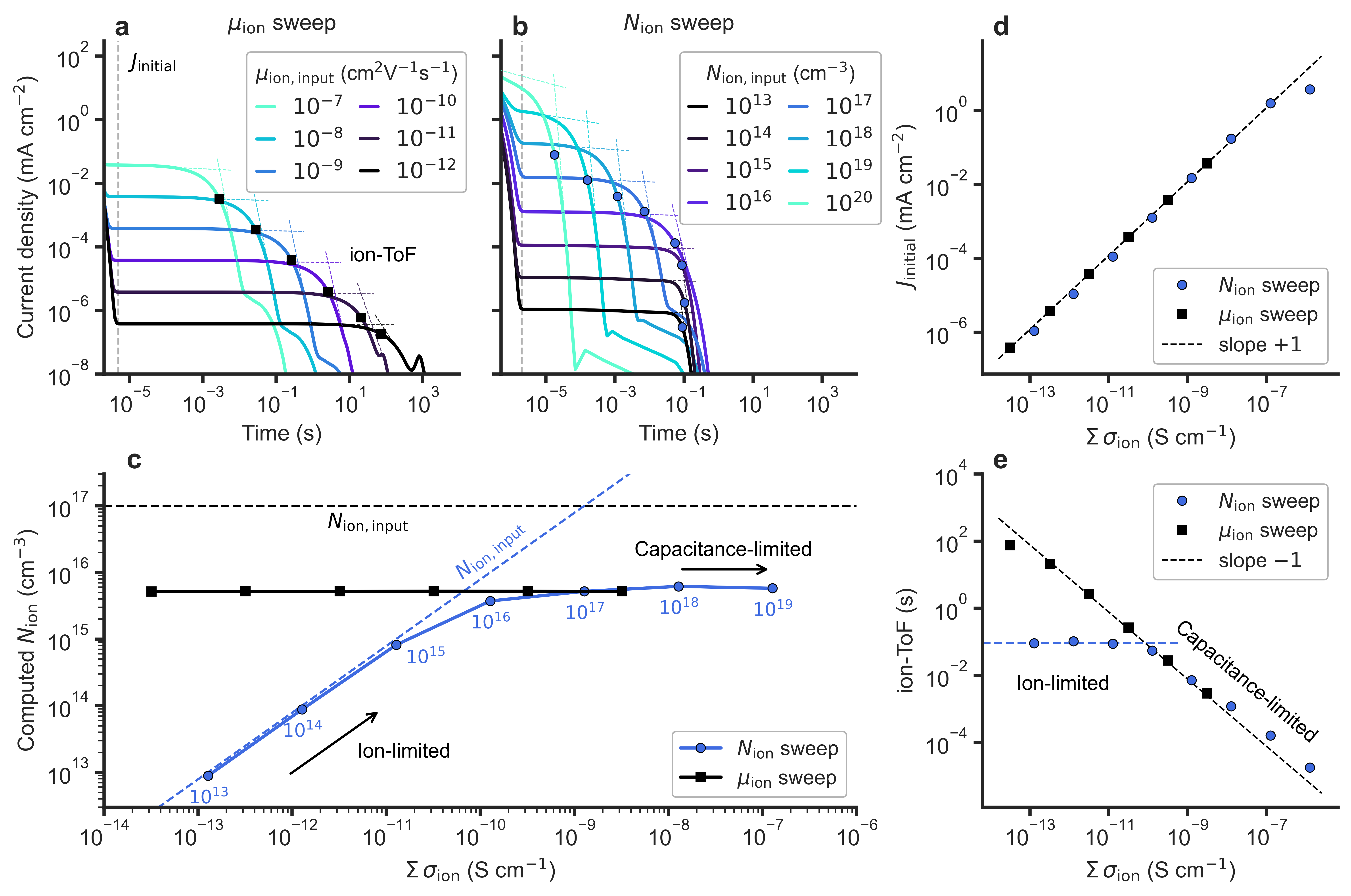}
  \caption{Impact of ionic properties on TIC (anion and cation properties varied simultaneously), showing simulated transient currents at 0\,V in the dark, after a preconditioning voltage ($V_{\rm pre}$) of 1.2\,V. The initial current spikes due to the fast electrode discharge are not fully shown. (a) Ion mobility $\mu_\mathrm{ion}$ sweep under constant $N_\mathrm{ion}$ ($10^{17}$\,cm$^{-3}$) and (b) input ion density $N_\mathrm{ion}$ sweep under constant $\mu_\mathrm{ion}$ ($4\times10^{-8}$\,cm$^{2}$\,V$^{-1}$\,s$^{-1}$). (c) Ion densities computed from the current transients in (a) and (b) from Eq.\,\eqref{eq:Nion} as a function of $\Sigma \sigma_{\mathrm{ion}}$, without $J_\mathrm{disp}$ correction and using constant $f_d=1$. (d) Initial current $J_\mathrm{initial}$ (extracted from the plateau region at 5\,$\mathrm{\mu}$s indicated by the grey dashed line in (a) and (b). (e) Time-of-flight (ion-ToF) for both parameter sweeps, obtained from the intersection of two linear fits to the TIC, as shown by the symbols in (a) and (b). The complete parameter set can be seen in Table\,S1.}
  \label{fig1}
\end{figure}

To assess the limitations of the TIC analysis, we first simulate its dependence on the ionic properties of the perovskite absorber and on the electronic properties of the TLs. For simplicity, initial drift-diffusion simulations are performed using a symmetrical model (in terms of parameters for electron and hole selectivity and transport to their respective contacts), with an equal density of mobile anions and cations confined to the perovskite layer (complete parameter list in Table\,S1). While we use the two-mobile-ion model as our baseline, the implications of the one-ion model are addressed later. For the voltage step, we simulate the TIC at 0\,V upon a switch from a preconditioning voltage of 1.2\,V, which corresponds to the work function (WF) difference between the metal electrodes, defined here as the built-in voltage ($V_\mathrm{bi}$).

Fig.\,\ref{fig1} shows the simulated transient currents, which originate from ion drift following the voltage switch, for a range of ionic properties. For now, we analyze only the total current ($J(t)$), as would be accessible in experiment. The effect of the displacement current $J_{\mathrm{disp}}$, which can partially cancel out the ion drift current in the measured signal (Eq.\,\ref{eq:Nion}),\cite{diethelm2025probing} is discussed later.

As shown in Fig.\,\ref{fig1}a-b, varying ion mobility ($\mu_{\mathrm{ion}}$) or density ($N_\mathrm{ion}$) can change the timescales similarly to varying temperature in experiment.\cite{torre2025ion,peterson2025reliable} 
Lower $\mu_{\mathrm{ion}}$ or $N_\mathrm{ion}$ reduces the initial ionic current consistently, while the duration of the current decay depends on the limiting regime, as discussed next. For the moment, we ignore $f_d$ and compute $N_\mathrm{ion}$ via Eq.\,\eqref{eq:Nion} with $f_d = 1$, as commonly done.\cite{peterson2025reliable} This would yield the correct ion density if both mobile cations and anions drifted on average $d_\mathrm{PSK}/2$ in opposite directions, equivalent to one species crossing the full layer. Further on, we will perform the necessary $f_d$ correction. 

Fig.\,\ref{fig1}c shows the computed $N_{\mathrm{ion}}$ as a function of the total ionic conductivity $\Sigma \sigma_{\mathrm{ion}}$, which in this case is simply $2 \cdot q \cdot \mu_{\mathrm{ion}} \cdot N_{\mathrm{ion}}$ since both ionic species (cations and anions) are considered mobile with the same mobility and density. For the mobility sweep, the computed $N_{\mathrm{ion}}$ remains independent of $\mu_{\mathrm{ion}}$, as desired. However, it remains below the input $N_\mathrm{ion}$ for the selected value (10$^{17}$ cm$^{-3}$). For the input $N_\mathrm{ion}$ sweep, the computed $N_\mathrm{ion}$ follows the input value 1:1 at low densities, but saturates above 10$^{16}$ cm$^{-3}$. Thus, in this device, the maximum measurable $N_{\mathrm{ion}}$ is limited to approximately 10$^{16}$ cm$^{-3}$.

Fig.\,\ref{fig1}d shows how the initial current density $J_\mathrm{initial}$ scales linearly with $\sigma_{\mathrm{ion}}$. The ion time-of-flight (ion-ToF), i.e. the timescale over which ions redistribute, scales inversely for both the $\mu_{\mathrm{ion}}$ sweep and $N_{\mathrm{ion}}$ sweep in the saturated region (Fig.\,\ref{fig1}e). However, it becomes constant at low input $N_{\mathrm{ion}}$.

According to Diethelm et al.\cite{diethelm2025probing}, the initial current due to ion drift in the perovskite bulk electric field can be described by
\begin{equation}
    J_\mathrm{initial} = \Sigma \sigma_{\mathrm{ion}} \cdot E_{\mathrm{bulk,initial}},
    \label{eq:Jion0}
\end{equation}
where $E_{\mathrm{bulk,initial}}$ is the bulk electric field after switching the applied bias (at $t_\mathrm{initial}$). If ions are uniformly distributed throughout the perovskite bulk during preconditioning and confined to the perovskite layer, no other sources of space charge are present, and the TLs are highly conductive, then $E_{\mathrm{bulk,initial}}$ at 0\,V could be approximated as $V_{\mathrm{bi}} / d_{\mathrm{PSK}}$. However, in practice a significant fraction of the built-in potential may drop across the TLs ($\Delta \phi_\mathrm{TL}$), so that the effective field driving ion drift at 0\,V becomes
\begin{equation}
    E_{\mathrm{bulk,initial}} = 
    \frac{V_{\mathrm{bi}} - \Sigma \Delta \phi_\mathrm{TL}}{d_{\mathrm{PSK}}} = 
    \frac{\Delta\phi_{\mathrm{PSK}}}{d_{\mathrm{PSK}}},
    \label{eq:Ebulk}
\end{equation}
where $\Delta\phi_{\mathrm{PSK}}$ is the part of $V_{\mathrm{bi}}$ dropped across the perovskite layer at $t_\mathrm{initial}$, when ions still retain their preconditioned distribution.

The reduction of the initial electric field can affect the quantified $\sigma_\mathrm{ion}$ from TIC using Eq.\,\eqref{eq:Jion0}. However, when computing $N_\mathrm{ion}$ from the integral of TIC (Eq.\,\eqref{eq:Nion}), the dominant limitation is set at the steady-state condition after the ionic current transient, not by $E_{\mathrm{bulk,initial}}$. The ionic current vanishes as soon as the capacitance per unit area set by the TLs ($C_\mathrm{TL}$) is fully charged, meaning that most of the potential drops at the TLs (i.e. $\Sigma \Delta \phi_\mathrm{TL} \to V_{\mathrm{bi}}$). As a consequence, the electric field in the perovskite bulk vanishes and the charge per unit area at each perovskite TL interface is roughly $Q_\mathrm{ion,max} = C_\mathrm{TL} \cdot V_\mathrm{bi}/2$ for our symmetric device. Thus, when the actual available ionic charge is lower than $C_\mathrm{TL} \cdot V_\mathrm{bi}/2$, the input ion density is fully quantified by TIC, and the measurement becomes ion-limited, as shown in Fig.\,\ref{fig1}c.
In this regime, the ion-ToF becomes independent of $N_\mathrm{ion}$ and scales with $\mu_\mathrm{ion}$, since it represents the ionic transit time $d_\mathrm{PSK}/(\mu_\mathrm{ion} \cdot E_{\mathrm{bulk,initial}})$. 
For large input $N_\mathrm{ion}$, the maximum measurable charge becomes limited by the TL capacitance (Fig.\,\ref{fig1}c). Consequently, the ion-ToF in the capacitance-limited regime reflects the $RC$ time constant for the ionic current to charge the TL capacitor, which explains the slope $-1$ dependence on $\sigma_{\mathrm{ion}}$ (Fig.\,\ref{fig1}e).

During degradation of PSCs, it is suspected and has been reported that one underlying cause can be a rising $N_{\mathrm{ion}}$.\cite{thiesbrummel2026ion} In the ion-limited case, this would be expected to increase $J_\mathrm{initial}$ (Fig.\,\ref{fig1}d) but leave the ion-ToF mostly unchanged (Fig.\,\ref{fig1}e). In contrast, in the capacitance limit, a rising $N_{\mathrm{ion}}$ in the perovskite bulk during aging would accelerate the ionic current transient (shortening the ion-ToF), as well as shift the peak $J$-$V$ hysteresis to faster scan rates (Fig.\,S1), both due to the rise in $\sigma_{\mathrm{ion}}$ (even if the computed $N_{\mathrm{ion}}$ from TIC would stay mostly constant). The peak hysteresis scan rate shows a similar dependence on the limiting regime as the ion-ToF, but the magnitude of $J$-$V$ hysteresis, which depends on both recombination properties and $N_{\mathrm{ion}}$,\cite{tress2017metal,torre2025ionic} tends to disappear in the ion-limited regime (Fig.\,S1), because $N_{\mathrm{ion}}$ is not high enough to screen the electric field and modify charge transport considerably. Timescale changes in TIC and $J$-$V$ hysteresis can therefore help to infer changes in $\sigma_{\mathrm{ion}}$ during degradation.\cite{kalasariya2026halide} If they are not consistently observed experimentally during aging,\cite{thiesbrummel2024ion,seid2025mitigating} either the device remains ion-limited, in which case ion-induced losses and $J$-$V$ hysteresis may remain small,\cite{torre2025ionic} or additional processes are at play, beyond a rising $N_{\mathrm{ion}}$ and the associated faster timescales. 

Furthermore, if the TIC preconditioning voltage $V_{\mathrm{pre}}$ differs from the voltage at which ions are compensated in the bulk, ionic accumulation at the interfaces may result in an initial bulk field that is partially screened or enhanced relative to the built-in field, depending on whether $V_{\mathrm{pre}}$ is below or above $V_{\mathrm{\min},Q}$. Preconditioning at $V_{\mathrm{\min},Q}$, which minimizes ionic accumulation at the interfaces, would yield the ionic distribution closest to an equivalent ion-free device,\cite{hart2024more,torre2025ionic} and would therefore be the condition required for Eq.\,\eqref{eq:Ebulk} to apply in terms of $V_{\mathrm{bi}}$. More generally, since $E_{\mathrm{bulk,initial}}$ depends on the applied voltage step ($V_{\mathrm{pre}}-V$),\cite{diethelm2025probing} Eq.\,\eqref{eq:Ebulk} can be applied at 0\,V with $V_{\mathrm{pre}}$ in place of $V_{\mathrm{bi}}$, and the potential partition taken relative to $V_{\mathrm{pre}}$. Such effects are discussed further below.

\begin{figure}[t]
\centering
  \includegraphics[width=\linewidth]{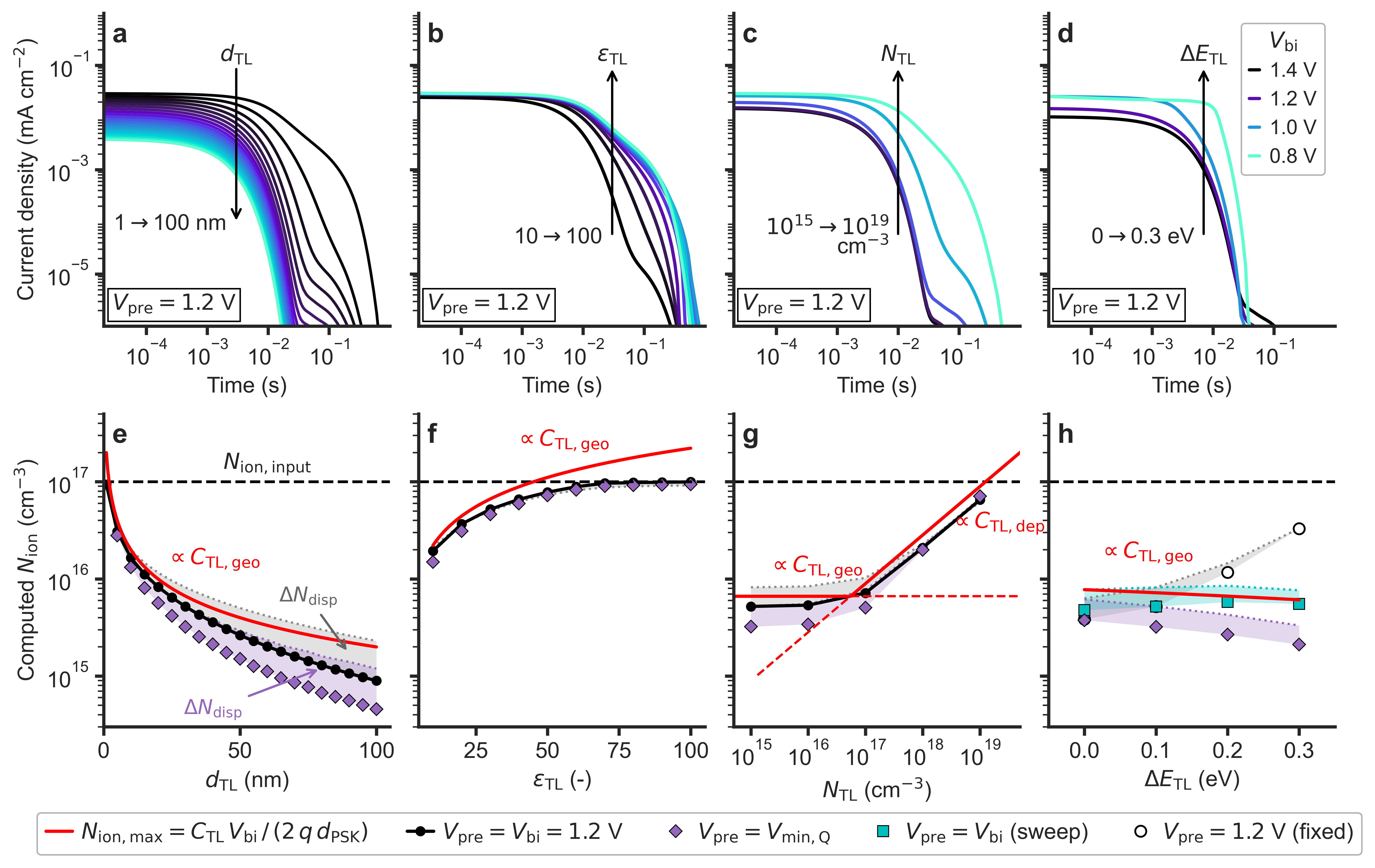}
  \caption{(a)-(d) Impact of TL properties on TIC, simulated at 0\,V in the dark, after a preconditioning voltage ($V_{\rm pre}$) of 1.2\,V. For the sweeps, both ETL and HTL properties are varied simultaneously, consisting of thickness ($d_{\rm TL}$), dielectric constant ($\varepsilon_{\rm TL}$), doping density ($N_{\rm TL}$), and the $V_{\rm bi}$ and energy-band offset ($\Delta E_{\rm TL}$) at the perovskite/TL interfaces. 
  (e)-(h) Computed $N_\mathrm{ion}$ (Eq.\,\eqref{eq:Nion}, with constant $f_d=1$), in which filled black circles correspond to preconditioning at $V_\mathrm{pre} = V_\mathrm{bi} = 1.2$\,V and grey diamonds to preconditioning at $V_{\mathrm{\min},Q}$. $\Delta N_\mathrm{disp}$ illustrates the reduction due to the opposing displacement current. In (h), where the sweep varies $V_\mathrm{bi}$ itself, cyan squares denote $V_\mathrm{pre}$ adjusted to follow $V_\mathrm{bi}$, and open circles a fixed $V_\mathrm{pre} = 1.2$\,V.
  The red solid curves show the capacitive limit $N_\mathrm{ion,max}$ (Eq.\,\eqref{eq: Nion_max}), evaluated with the geometric capacitance ($C_\mathrm{TL,geo}$, Eq.\,\eqref{eq: C_geo}) or, where doping dominates, the effective capacitance $C_\mathrm{TL,eff}$ ($\propto C_\mathrm{TL,dep}$, Eq.\,\eqref{eq: C_dep}).
  The simulation parameters are summarized in Table\,S1.}
  \label{fig2}
\end{figure}

To highlight the limiting factors for the quantification of mobile ions using TIC, we simulated the effect of varied TL parameters, including thickness ($d_{\rm TL}$), dielectric constant ($\varepsilon_{\rm TL}$), doping density ($N_{\rm TL}$), and the energy-band alignment between the perovskite and TLs ($\Delta E_{\rm TL}$), as illustrated in Fig.\,\ref{fig2} (both HTL and ETL parameters are varied simultaneously). 

The parameter sweeps are designed so that different contributions to $C_{\mathrm{TL}}$ dominate in turn. In general, the charge per unit area accumulated on the TL side of the PSK/TL interface at a potential drop $\Delta\phi_{\mathrm{TL}}$ is
\begin{equation}
    Q_{\mathrm{TL}} \;=\; \int_{0}^{\Delta\phi_{\mathrm{TL}}} C_{\mathrm{TL}}(V)\,\mathrm{d}V,
    \label{eq: Q_int}
\end{equation}
which reduces to $Q_{\mathrm{TL}} = C_{\mathrm{TL}}\,\Delta\phi_{\mathrm{TL}}$ only when $C_{\mathrm{TL}}$ is voltage-independent.

For the $d_{\rm TL}$ and $\varepsilon_{\rm TL}$ sweeps, the TLs are kept undoped, such that $C_{\mathrm{TL}}$ is set by the (voltage-independent) geometric capacitance $C_{\mathrm{TL,geo}}$:
\begin{equation}
    C_{\mathrm{TL,geo}} \;=\; \frac{\varepsilon_0\,\varepsilon_{\mathrm{TL}}}{w},
  \label{eq: C_geo}
\end{equation}
where the relevant width $w$ equals the full TL thickness $d_{\mathrm{TL}}$, and Eq.\,\ref{eq: Q_int} becomes $Q_{\mathrm{TL,geo}} = C_{\mathrm{TL,geo}}\,\Delta\phi_{\mathrm{TL}}$. With sufficient doping ($N_{\rm TL}$), the depletion region instead governs $C_{\mathrm{TL}}$ with a width smaller than $d_{\mathrm{TL}}$ (Fig.\,S2):\cite{sze2021physics}
\begin{equation}
    w_{\mathrm{TL,dep}} \;=\; \sqrt{\frac{2\,\varepsilon_0\,\varepsilon_{\mathrm{TL}}\,\Delta\phi_{\mathrm{TL}}}{q\,N_{\mathrm{TL}}}}.
    \label{eq: W_dep}
\end{equation}
Because the corresponding differential capacitance is voltage-dependent, $C_{\mathrm{TL,dep}}(V)=\sqrt{q\,\varepsilon_0\,\varepsilon_{\mathrm{TL}}\,N_{\mathrm{TL}}/(2V)}$, Eq.\,\ref{eq: Q_int} yields
\begin{equation}
    Q_{\mathrm{TL,dep}} \;=\; q\,N_{\mathrm{TL}}\,w_{\mathrm{TL,dep}} \;=\; \sqrt{2\,q\,\varepsilon_0\,\varepsilon_{\mathrm{TL}}\,N_{\mathrm{TL}}\,\Delta\phi_{\mathrm{TL}}},
    \label{eq: Q_dep}
\end{equation}
which can be used to define an effective capacitance
\begin{equation}
    C_{\mathrm{TL,eff}} \;\equiv\; \frac{Q_{\mathrm{TL,dep}}}{\Delta\phi_{\mathrm{TL}}} \;=\; 2\,C_{\mathrm{TL,dep}}(\Delta\phi_{\mathrm{TL}}).
    \label{eq: C_dep}
\end{equation}

In the limit $\Sigma\,\Delta\phi_\mathrm{TL} \to V_{\mathrm{bi}}$, the ionic charge $Q = \int_0^\infty J(t)\,\mathrm{d}t$ (Eq.\,\ref{eq:Nion}, assuming $J_{\mathrm{ion}} \gg J_{\mathrm{disp}}$) accumulated at each interface is limited by $Q_{\mathrm{TL,geo}}$ or $Q_{\mathrm{TL,dep}}$ (Eq.\,\ref{eq: Q_dep}), whichever is higher. For the symmetric device with $\Delta\phi_{\mathrm{TL}} = V_{\mathrm{bi}}/2$, the maximum measurable $N_{\mathrm{ion}}$ is set by the charge each TL capacitor can accommodate:

\begin{equation}
    N_{\mathrm{ion,max}} \;=\; \frac{C_{\mathrm{TL}}\,V_{\mathrm{bi}}}{2\,q\,d_{\mathrm{PSK}}}.
    \label{eq: Nion_max}
\end{equation}

When the available ion density exceeds this capacitive limit ($N_{\mathrm{ion}} > N_{\mathrm{ion,max}}$), each species shifts by less than $d_{\mathrm{PSK}}/2$), so that $f_d < 1$ for two mobile ions (or $f_d < 0.5$ for one mobile ion). For asymmetric transport layers ($C_{\mathrm{HTL}} \neq C_{\mathrm{ETL}}$), the two TLs form a series connection and the measurable charge is the change on the external electrodes, giving

\begin{equation}
    N_{\mathrm{ion,max}} = \frac{C_{\mathrm{HTL}}\,C_{\mathrm{ETL}}\,V_{\mathrm{bi}}}{(C_{\mathrm{HTL}} + C_{\mathrm{ETL}})\,q\,d_{\mathrm{PSK}}},
    \label{eq:Nion_max_asym}
\end{equation}

which reduces to Eq.~\ref{eq: Nion_max} above for $C_{\mathrm{HTL}} = C_{\mathrm{ETL}} = C_{\mathrm{TL}}$.

As shown in Fig.\,\ref{fig2}a\&e, increasing $d_{\rm TL}$ leads to a significant reduction in both $J_\mathrm{initial}$ and the ion-ToF, resulting in a decrease of the computed $N_{\rm ion}$ in agreement with the capacitive-limit imposed by $C_{\mathrm{TL,geo}}$. A similar dependence is obtained with varying $\varepsilon_{\rm TL}$ (Fig.\,\ref{fig2}b\&f), which also modifies $C_{\mathrm{TL,geo}}$ through Eq.\,\eqref{eq: C_geo}. Increasing the doping density $N_{\mathrm{TL}}$ (Fig.\,\ref{fig2}c\&g) raises $C_{\mathrm{TL}}$ above the geometric value via the depletion capacitance (Eq.\,\eqref{eq: C_dep}), which increases the computed $N_{\mathrm{ion}}$ proportionally to $\sqrt{N_{\mathrm{TL}}}$ in this case. 

To some extent, the computed $N_{\mathrm{ion}}$ may also depend on the initial bulk electric field, and in turn on how far the preconditioning voltage $V_{\mathrm{pre}}$ is from $V_{\mathrm{\min},Q}$, as shown in Fig.\,\ref{fig2}e-h, where different markers correspond to $V_{\mathrm{pre}}=1.2$\,V and $V_{\mathrm{\min},Q}$. The latter ($V_{\mathrm{\min},Q}$) can fall below $V_{\mathrm{bi}}$ for thick or undoped TLs.\cite{hart2024more} Thus, the $V_{\mathrm{pre}}$ used may also affect the trends obtained, which may be important when trying to compare ionic parameters across devices with different TLs.\cite{torre2025ionic} To illustrate a relevant case, Fig.\,\ref{fig2}d shows a variation of the energetic alignment of the device, varying both $V_{\mathrm{bi}}$ and the energy band offsets at the perovskite/TL interfaces simultaneously ($q V_{\mathrm{bi}}=1.4\,\mathrm{eV}-2\Delta E_{\mathrm{TL}}$). The lower $V_{\mathrm{bi}}$ and higher injection barriers to perovskite under forward bias, can lead to varied initial ionic accumulation, if the same 1.2\,V precondition is used. The computed $N_{\mathrm{ion}}$ for the energy-band alignment sweep only remains close to the $C_{\mathrm{TL,geo}}$-limit when $V_{\mathrm{pre}}$ is adjusted such that ions remain mostly uniformly distributed at the initial state. With fixed preconditioning, the computed $N_\mathrm{ion}$ can increase beyond the simple capacitive limit, as the accumulation layers invert polarity if $V_{\mathrm{pre}}$ is high above $V_{\mathrm{\min},Q}$, effectively changing the measurable charge. 
In Fig.\,\ref{fig2}, when the TLs have a high $\varepsilon_{\rm TL}$ ($\geq 70$) or high $N_{\rm TL}$ ($\geq 10^{19}$ cm$^{-3}$), the displaced $N_{\mathrm{ion}}$ can reach the ion-limited value at fixed TL thickness ($d_{\rm TL}$ = 30\,nm in Fig.\,\ref{fig2}b-d), since the true $N_{\mathrm{ion}}$ becomes measurable when it is below the capacitive limit.

As discussed above, two key mechanisms can affect TIC; the initial bulk electric field at the precondition, and the final accumulation of ions at the TL interfaces. For $V_{\mathrm{pre}}$ below $V_{\mathrm{min},Q}$ the initial accumulation can reduce $E_{\mathrm{bulk,initial}}$ and partially screen $V_{\mathrm{bi}}$, only slightly reducing the computed $N_{\mathrm{ion}}$, such that the initial field (Eq.\,\eqref{eq:Ebulk}) and capacitive limit (Eq.\,\ref{eq: Nion_max}) expressions remain valid when evaluated at the applied $V_{\mathrm{pre}}$. For $V_{\mathrm{pre}}$ above $V_{\mathrm{min},Q}$ the accumulation layers switch polarity, and the initial discharge process can then add considerably to the current transient; at high forward bias, due to charge injection from the electrodes, the TLs may also become increasingly conductive rather than purely geometric capacitors, so that the resulting $N_{\mathrm{ion}}$ from TIC can even exceed the capacitance-limited estimate (e.g. Fig.\,\ref{fig2}h). This effect is addressed further in Section\,\ref{sec:above-limit}.

\begin{figure}[t]
\centering
  \includegraphics[width=\linewidth]{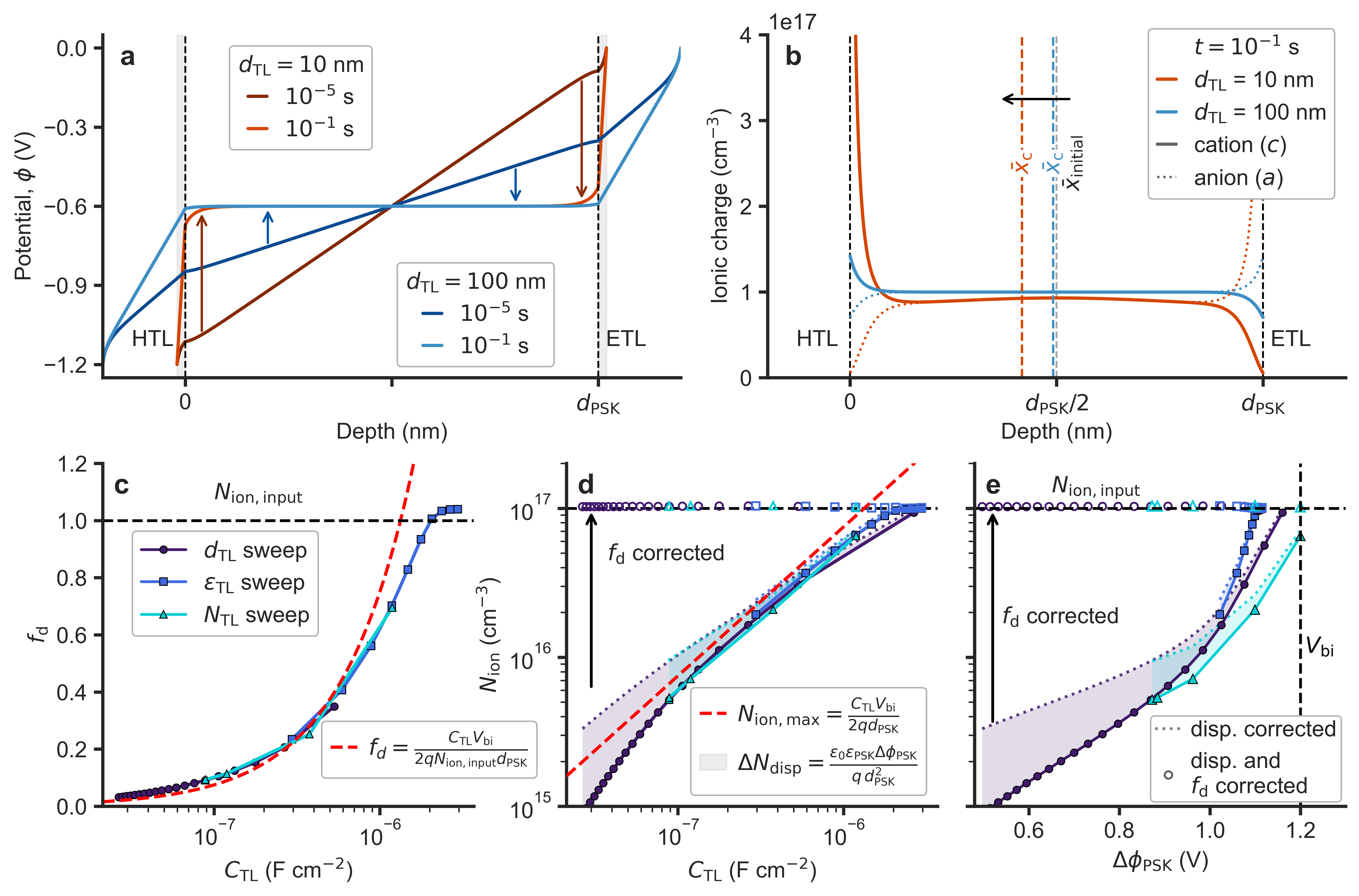}
  \caption{(a) Potential $\phi$ across the stack and (b) ionic charge profiles (cations $c$ and anions $a$) for 10\,nm and 100\,nm TL thickness ($d_{\mathrm{TL}}$), showing the time before ions respond ($10^{-5}$\,s), and after ionic redistribution ($10^{-1}$\,s). The arrows indicate the evolution with time. 
  The average displacement of the cation distribution is shown as the dashed lines (the anion one, not shown, is symmetric in the opposite direction from $d_{\mathrm{PSK}}/2$). 
  (c) Correction factor $f_d$, derived from the average ionic displacement via Eq.\,\eqref{eq:CoM}, as a function of TL capacitance ($C_{\mathrm{TL}}$). The capacitive limit (red dashed line) is plotted as $f_d =N_{\mathrm{ion,max}}/N_{\mathrm{ion,input}}$.
  (d)-(e) Computed $N_{\mathrm{ion}}$ from the total current integral (filled symbols) and after both the displacement current (disp.) and $f_d$ corrections (open symbols), versus (d) $C_{\mathrm{TL}}$ and (e) initial $\Delta\phi_{\mathrm{PSK}}$. In (d), the red dashed line represents the capacitive limit $N_{\mathrm{ion,max}}$ (Eq.\,\ref{eq: Nion_max}).}
  \label{fig3}
\end{figure}

To illustrate the physical mechanisms governing the observed trends, Fig.\,\ref{fig3} shows the simulated electrical profiles for varied $d_{\rm TL}$, corresponding to the TIC results in Fig.\,\ref{fig2} for fixed 1.2\,V preconditioning and switching to 0\,V. Fig.\,\ref{fig3}a compares the potential drops across the PSC for 10- versus 100\,nm-$d_{\rm TL}$. Immediately after the voltage switch ($t=10^{-5}$\,s), before ions have redistributed, the potential drop in perovskite is higher in the device with thin TLs. Consequently, the 10\,nm-$d_{\rm TL}$ device exhibits a higher ion drift current $J_\mathrm{initial}$ (Eq.\,\eqref{eq:Jion0}-\eqref{eq:Ebulk}) for the same $V_{\mathrm{pre}}$. At $t=10^{-1}$\,s, after ions have redistributed, most of the potential drops across the TLs in both cases ($\Sigma \Delta \phi_\mathrm{TL} \to V_{\mathrm{bi}}$), indicative of the fact that the TL capacitances are fully charged. Fig.\,\ref{fig3}b shows the final ionic distributions at equilibrium and the displacement of the cation density centroids $\Delta \bar{x}_c$. For anions the displacement would in this case look identical but in the opposite direction. For simplicity, the $d_{\rm PSK}/2$ line is shown as $\bar{x}_{\rm initial}$ in Fig.\,\ref{fig3}b, which would be the state under $V_{\mathrm{\min},Q}$-preconditioning. For the current simulation settings, the fixed 1.2\,V precondition varies $\bar{x}_{\rm initial}$, depending on the initial level of ionic accumulation. The change in $C_\mathrm{TL}$ with varying $d_\mathrm{TL}$ leads to a different average displacement of the ionic distributions. As demonstrated next, taking this into account can be used to correct the computed $N_{\mathrm{ion}}$ from TIC.

\subsection{$f_d$ correction and TL-free extrapolation}

The correction factor $f_d$ obtained via Eq.\,\eqref{eq:CoM} (using the actual $\bar{x}_{\rm initial}$ of the initial ionic distributions under 1.2\,V preconditioning) is shown in Fig.\,\ref{fig3}c as a function of $C_{\mathrm{TL}}$ (Eq.\,\eqref{eq: C_geo}\&\eqref{eq: C_dep}), combining all parameter sweeps from Fig.\,\ref{fig2}a-c. The calculated $f_d$ values mostly collapse onto a single trend, confirming that the $N_{\mathrm{ion}}$ computed via Eq.\,\eqref{eq:Nion} with constant $f_d=1$ is dominated by the capacitance of the TLs. As $C_\mathrm{TL}$ decreases, fewer ions are displaced to the interfaces (Fig.\,\ref{fig3}b\&Fig.\,S3), and the average ionic displacement decreases accordingly. $f_d$ therefore reduces with $C_\mathrm{TL}$ (Fig.\,\ref{fig3}c\&Fig.\,S4), meaning that the fraction of the available ion density that accumulates at the perovskite/TL interfaces is limited by $C_\mathrm{TL}$.

Consequently, $f_d$ can be expressed analytically as an ideal correction based on the fraction of the maximum measurable charge in the capacitance-limited regime (Eq.\,\eqref{eq: Nion_max}), i.e.\ $N_{\mathrm{ion,max}}/N_\mathrm{ion,input}$, shown as the red dashed line in Fig.\,\ref{fig3}c, which exhibits strong agreement with the simulated data. 

The displacement current ($J_\mathrm{disp}$), which has the opposite sign to the ionic current on the relevant timescales (Fig.\,S5), reduces the charge density obtained by integrating TIC by an amount $\Delta N_\mathrm{disp}$; its contribution is relatively small but becomes considerable at lower $C_\mathrm{TL}$ (Fig.\,\ref{fig2} and Fig.\,S6). Since $J_\mathrm{disp} \propto \mathrm{d}E_\mathrm{bulk}/\mathrm{d}t$, its time integral reduces to the change in bulk field itself. Assuming the bulk field is fully screened in the final state, this gives $\int J_\mathrm{disp}(t)\,\mathrm{d}t = \varepsilon_0\varepsilon_\mathrm{PSK}\,E_\mathrm{bulk,initial}$, with $E_\mathrm{bulk,initial}$ from Eq.\,\eqref{eq:Ebulk}. In Eq.\,\ref{eq:Nion}, we can add this term to the raw $\int J(t)\,\mathrm{d}t$ to obtain the displacement current-corrected ionic charge per unit area. Equivalently, for the ionic charge density directly, $\Delta N_\mathrm{disp}=\int J_\mathrm{disp}(t)\,\mathrm{d}t/(q\,d_\mathrm{PSK})$ can be added to the computed $N_\mathrm{ion}$, as shown by the dotted lines in Fig.\,\ref{fig3}d-e for all parameter sweeps. Applying the further $f_d$ correction of Eq.\,\ref{eq:Nion} on top of this, $(N_\mathrm{ion}+\Delta N_\mathrm{disp})/f_d$ (open markers), reveals the true input $N_\mathrm{ion}$ throughout the capacitance-limited regime.

Fig.\,\ref{fig3}e shows the computed $N_{\mathrm{ion}}$ versus the amount of $V_{\mathrm{bi}}$ that drops in the perovskite layer ($\Delta\phi_\mathrm{PSK}$) in the initial state of TIC, at 0\,V before ionic redistribution. Because the $f_d$ correction is itself dependent on $N_{\mathrm{ion}}$, a linear rescaling to correct for the difference in initial $\Delta\phi_\mathrm{PSK}$ alone (i.e., $N_\mathrm{ion} \cdot (\Delta\phi_\mathrm{PSK}/V_\mathrm{bi})^{-1}$), as discussed in ref.\,\cite{schmidt2025quantification}, is insufficient to recover the true ionic charge in this case (see also Fig.\,S7). The trends obtained are strongly dominated by how much charge can be stored in the TLs in equilibrium at 0\,V. Hence conceptually it is more appropriate to refer to the saturation of the measurable ionic charge in TIC as a capacitance-limit, rather than an electric field-limit; even if $V_{\mathrm{pre}}$ is adjusted to result in the same initial field ($E_{\mathrm{bulk,initial}}$) for different $\Delta\phi_\mathrm{PSK}$ due to changes in $C_{\mathrm{TL}}$, the computed $N_\mathrm{ion}$ still scales with $C_{\mathrm{TL}} \propto 1/d_{\mathrm{TL}}$ (Fig.\,S8). Thus, the key to accessing the ion-limited regime, in which $f_d$ remains high and Eq.\,\eqref{eq:Nion} reflects the total intrinsic ion density (without correction), is to increase $C_\mathrm{TL}$ such that it is not limiting.

To further probe the fundamental limits of TIC in simulations, we now look at a TL-free device, in which the perovskite layer is contacted directly by two metal electrodes, whose WF are still set $0.2$\,eV below the conduction band and above the valence band at the electron and hole contacts, respectively ($V_\mathrm{bi} = 1.2$\,V).

\begin{figure}[t]
\centering
  \includegraphics[width=\linewidth]{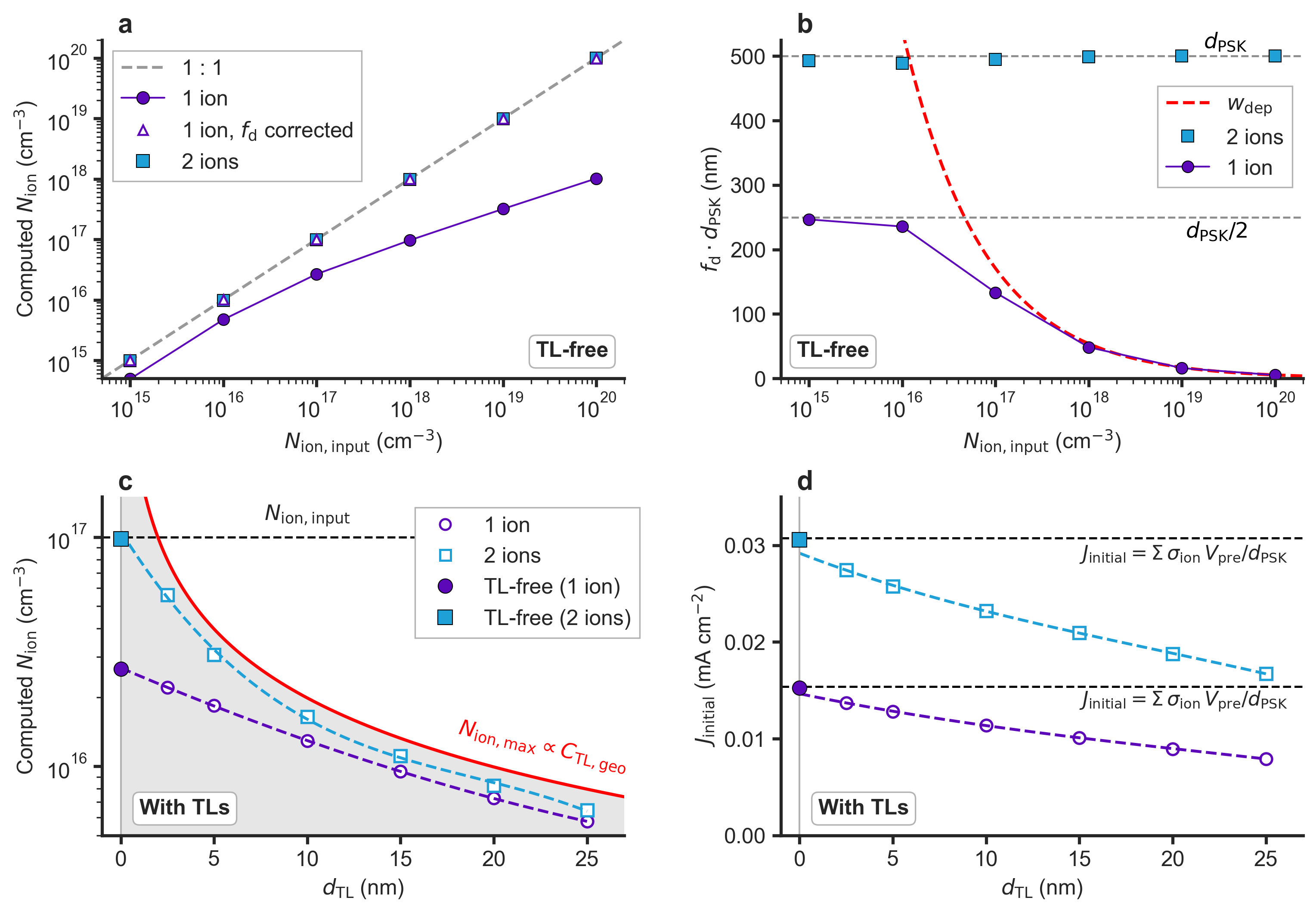}
  \caption{(a) Computed $N_{\mathrm{ion}}$ (Eq.\,\eqref{eq:Nion}) from TL-free simulations ($d_{\mathrm{TL}}$=0) for mobile cations only (1 ion, with a uniformly distributed static negative countercharge) and for both mobile anions and cations (2 ions), varying the input ion density. A fixed geometric prefactor $f_d = 1$ is used, except for the $f_d$-corrected case (Eq.\,\ref{eq: W_dep_PSK}). (b) Average ionic displacement ($f_d \cdot d_{\mathrm{PSK}}$ from Eq.\,\eqref{eq:CoM}) for the different models, from the initial state at $V_\mathrm{pre} = V_\mathrm{bi}$ to equilibrium at 0\,V, and perovskite depletion width ($w_{\mathrm{dep}}$) approximation using Eq.\,\eqref{eq: W_dep_PSK} for varied $N_{\mathrm{ion}}$. (c) Computed $N_{\mathrm{ion}}$ assuming $f_d = 1$ and (d) $J_\mathrm{initial}$ for the $d_{\mathrm{TL}}$ sweep for 1 and 2 ion models, showing the trend fit (dashed line) used to extrapolate towards $d_{\mathrm{TL}}$=0. Filled markers show the simulated TL-free values in (a) for $10^{17}$ input $N_{\mathrm{ion}}$, without correction. The red curve in (c) is the capacitive limit $N_{\mathrm{ion,max}}\propto C_{\mathrm{TL,geo}}$.}
  \label{fig4}
\end{figure}

Fig.\,\ref{fig4}a shows the $N_\mathrm{ion}$ computed via Eq.\,\eqref{eq:Nion} with $f_d=1$, from simulated TIC transients at $0$\,V (again switched from $V_\mathrm{pre} = V_\mathrm{bi}$), as a function of the input $N_\mathrm{ion}$. In these TL-free devices, the limitation depends critically on the assumed ionic mobility model. As shown in Fig.\,\ref{fig4}a, the TL-free two-ion model follows the input $N_\mathrm{ion}$ along a 1:1 line across the full tested range, since both cations and anions migrate to opposite interfaces, fully depleting the bulk so that the average position of each species consistently shifts by $d_\mathrm{PSK}/2$. In this case, the situation remains ion-limited at all tested ion densities. 
In contrast, the one-ion model, which includes a constant immobile negative background charge of input $N_\mathrm{ion}$, analogous to ionized acceptors in a p-type semiconductor, may be limited even in the TL-free case.\cite{diekmann2023determination} 
Cation depletion at one interface exposes the background charge up to a maximum charge density of $-q N_\mathrm{ion}$, while cation accumulation at the other interface can form a much thinner positive layer (Fig.\,S9). The width of the ionic depletion region in PSK ($w_{\mathrm{dep}}$) follows the semiconductor depletion approximation\cite{garcia2025ionic}

\begin{equation}
\begin{split}
    w_{\mathrm{dep}}=\sqrt{\frac{2\,\varepsilon_0 \varepsilon_{\mathrm{PSK}} V_{\mathrm{bi}}}{q N_{\mathrm{ion}}}}.
\end{split}
\label{eq: W_dep_PSK}
\end{equation}

With increasing input $N_\mathrm{ion}$, the higher ionic space-charge density localizes the potential drop within a thinner interfacial depletion region; the remaining perovskite bulk therefore becomes nearly field-free and ions there do not redistribute further. The average ionic displacement then follows $w_{\mathrm{dep}}$ directly (Fig.\,\ref{fig4}b); the cations within the depletion region traverse nearly the full perovskite thickness to the accumulation layer, while the stationary majority does not contribute, so the average displacement is approximately the participating fraction ($w_{\mathrm{dep}}/d_\mathrm{PSK}$) times $d_\mathrm{PSK}$, i.e. $\Delta \bar{x}=f_d \cdot d_\mathrm{PSK} \approx w_\mathrm{dep}$.

Only at low $N_\mathrm{ion}$, where $w_{\mathrm{dep}} > d_\mathrm{PSK}/2$, more than half of the perovskite layer becomes depleted (Fig.\,S10) and the average displacement saturates at $d_\mathrm{PSK}/2$ (i.e. $f_d=0.5$). Applying the $f_d$ correction with average ionic displacement from Eq.\,\eqref{eq:CoM} brings the one-ion case onto the 1:1 line (Fig.\,\ref{fig4}a), demonstrating that the two models are consistent and TIC measurements can reflect the input ion density when the average ionic displacement is accounted for. Moreover, since the decrease of average ionic displacement below $d_\mathrm{PSK}/2$ with increasing $N_\mathrm{ion}$ follows the solution for $w_{\mathrm{dep}}$ (Fig.\,\ref{fig4}b), the computed one-ion model TL-free $N_\mathrm{ion}$ values can equivalently be corrected via $f_d \approx w_{\mathrm{dep}}/d_\mathrm{PSK}$, without requiring simulated ion distributions. 

\begin{table}[t]
    \centering
    \renewcommand\arraystretch{1.4}
    \caption{Comparison of input and extrapolated parameters in the simulation. $N_\mathrm{ion}$ is given per species (equal $N_\mathrm{ion}$ for anions and cations). In one-ion model, anions are homogeneously distributed and static.}
    \resizebox{\textwidth}{!}{%
    \begin{tabular}{cccccc}
    \toprule
        &  Input $\Sigma \sigma_\mathrm{ion}$ (S cm$^{-1}$) & Extrapolated $\Sigma \sigma_\mathrm{ion}$ (S cm$^{-1}$) &   Input $N_\mathrm{ion}$ (cm$^{-3}$) & Extrapolated $N_\mathrm{ion}$ (cm$^{-3}$) & $f_d$-corrected $N_\mathrm{ion}$ (cm$^{-3}$) \\
    \midrule
        \textbf{Two-ion} & $2\times6.41\times10^{-10}$ & $2\times6.08\times10^{-10}$ & $1\times10^{17}$ & $1.07\times10^{17}$ & - \\
        \textbf{One-ion} &  $6.41\times10^{-10}$ & $6.10\times10^{-10}$ & $1\times10^{17}$ & $2.70\times10^{16}$ & $1.01\times10^{17}$ \\    
    \bottomrule
    \end{tabular}%
    }
    \label{table: fitted results of figure 4}
\end{table}

For the TL-free devices, uncertainty enters if $V_\mathrm{pre}$ would differ greatly from $V_{\mathrm{bi}}$ (Fig.\,S11). However, if this difference can be assumed small, both the two-ion and one-ion assumptions allow an exact quantification of $N_\mathrm{ion}$ using TIC on TL-free devices. Therefore, we propose a protocol based on varying $d_{\mathrm{TL}}$ to quantify ionic properties in the capacitance-dominated (but not fully limited) regime by extrapolating towards the TL-free reference (where $\Delta\phi_\mathrm{TL} \to 0$). Fig.\,\ref{fig4}c shows how the $N_{\mathrm{ion}}$ obtained from the $d_{\mathrm{TL}}$-sweep simulations (with Eq.\,\eqref{eq:Nion} and $f_d=1$) follows a monotonic trend with TL thickness. The trend can be extrapolated to $d_{\mathrm{TL}}=0$, landing at the simulated TL-free value, albeit with the one-ion TL-free $N_{\mathrm{ion}}$ differing from the input value (10$^{17}$\,cm$^{-3}$) due to $f_d < 1$, as discussed above. We note that, depending on TL properties and ionic model assumed, the deviation from the capacitive limit may become less obvious and the extrapolation more challenging for high $N_\mathrm{ion}$ values (e.g., 10$^{19}$\,cm$^{-3}$), where the trend is steep at small $d_\mathrm{TL}$ (Fig.\,S12-S13).

Under the two-ion assumption, the extrapolation to TL-free directly yields the true $N_{\mathrm{ion}}$, since cations and anions each drift $d_{\mathrm{PSK}}/2$ in opposite directions, giving $f_d=1$. For a single mobile species, $f_d$ can be estimated as $w_{\mathrm{dep}}/d_{\mathrm{PSK}}$ when the depletion width is narrower than $d_{\mathrm{PSK}}/2$, or otherwise taken as $0.5$ (full depletion of a uniform initial distribution). Once the true $N_{\mathrm{ion}}$ is established from the TL-free extrapolation and the chosen ionic model, $f_d$ at any TL thickness can be approximated as the fraction of measured charge, to estimate how the average ionic displacement decreases in the capacitance-limited regime.

The conductivity $\sigma_{\mathrm{ion}}$ is similarly obtained by extrapolating the $d_{\mathrm{TL}}$ trend of the initial ionic current $J_\mathrm{initial}$, as shown in Fig.\,\ref{fig4}d. Since the extrapolation towards the TL-free limit corresponds to the case where $\Delta\phi_{\mathrm{PSK}}=V_{\mathrm{bi}}$, the ionic conductivity can be obtained directly as $\Sigma \sigma_{\mathrm{ion}} = J_\mathrm{initial}(d_\mathrm{TL}=0) \cdot (V_{\mathrm{bi}}/d_{\mathrm{PSK}})^{-1}$ using Eq.\,\eqref{eq:Jion0}. Provided $V_{\mathrm{pre}}$ remains not too high above $V_{\mathrm{min},Q}$, $V_{\mathrm{bi}}$ can be substituted for $V_{\mathrm{pre}}$ in this expression without significant error (Fig.\,S14). The extrapolated conductivities agree with the simulation inputs to within a few percent, in this case for both the one- and two-ion models (Table\,\ref{table: fitted results of figure 4}), validating the extrapolation approach. The $J_\mathrm{initial}$ extrapolation is straightforward across all tested $N_\mathrm{ion}$, since it is independent of $f_d$ (Fig.\,S12-S13) and only decreases due to the reduction of the initial electric field with $d_{\mathrm{TL}}$ (Eq.\,\eqref{eq:Ebulk}). We note that although the relative contribution of displacement current grows with $d_{\mathrm{TL}}$ (shown in Fig.\,S5), the extrapolation of the trend in initial measured current ($J_\mathrm{initial}$) still allows to obtain the true $\sigma_{\mathrm{ion}}$.

\subsection{Experimental verification}

\subsubsection{TL thickness variation of p-i-n devices}

\begin{figure}[t!]
\centering
  \includegraphics[width=\linewidth]{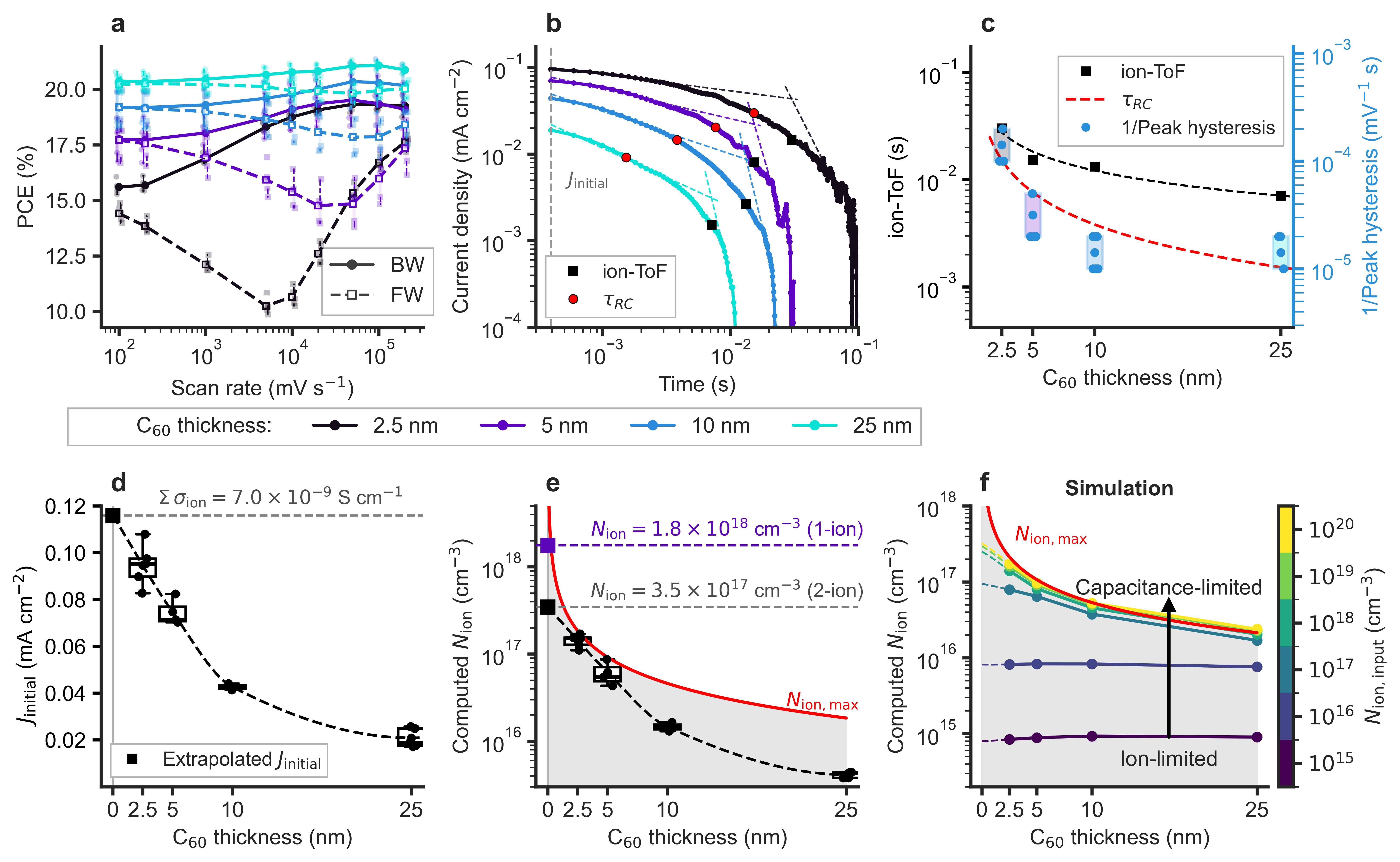}
  \caption{Experimental validation and simulation of TL-dependent ionic quantification for p-i-n devices. 
  (a) Statistical power conversion efficiency (PCE) as a function of scan rate, from $V_{\rm OC}$ precondition.
  (b) Measured TIC for different $\text{C}_{60}$ thicknesses. 
  (c) Ion time-of-flight (Ion-ToF) from (b) together with the inverse of the scan rate of maximum hysteresis from (a) (right axis). The red dashed curve shows the expected RC time constant, $\tau_\mathrm{RC} = R_{\mathrm{ion}} \cdot C_{\mathrm{TL,geo}} \propto 1/d_{\mathrm{ETL}}$.
  (d) Initial current $J_\mathrm{initial}$ with a trend fit (black dashed line) used to extrapolate to the TL-free limit. 
  (e) Statistical $N_\mathrm{ion}$ from integrating the TIC, with the trend fit (black dashed line) and the corresponding extrapolated TL-free $N_\mathrm{ion}$ (squares) with $f_d=1$ (two-ion) and $f_d=w_\mathrm{dep}/d_\mathrm{PSK}$ (one-ion), calculated using Eq.\,\eqref{eq: W_dep_PSK}.
  (f) Simulated results for a p-i-n model across various $\text{C}_{60}$ thicknesses and input $N_\mathrm{ion}$. The maximum measurable charge, evaluated using the $C_\mathrm{TL,geo}$ of C$_{60}$ (neglecting the SAM), is shown as the red line in (e) and (f). The simulations utilize the parameters detailed in Table\,S2.}
  \label{fig5}
\end{figure}

To verify the impact of TL properties on TIC as shown above with drift-diffusion simulations, we conducted a comparative study between measurements and simulations across both p-i-n and n-i-p PSC architectures. Within each fabrication series, only one TL parameter was intentionally varied, while all other layers and processing conditions were kept unchanged. For the p-i-n configuration (ITO / 4PADCB / Cs$_{0.05}$\allowbreak (FA$_{0.95}$\allowbreak MA$_{0.05}$)$_{0.95}$\allowbreak Pb(I$_{0.95}$\allowbreak Br$_{0.05}$)$_3$ (1.56\,eV) / C$_{60}$ / BCP / Cu), the $\text{C}_{60}$ ETL thickness was varied. Cross-sectional scanning electron microscopy (SEM) confirms a perovskite absorber thickness ($d_\mathrm{PSK}$) of approximately 600\,nm across all devices, with different $\text{C}_{60}$ thicknesses (Fig.\,S15).

Fig.\,\ref{fig5}a shows the PCE hysteresis across all C$_{60}$ thicknesses ($d_\mathrm{ETL}$). From a $V_\mathrm{OC}$ precondition, all devices reach a reasonably high PCE at high scan rates, since measuring faster than the ionic response prevents $E_\mathrm{bulk}$ screening. At intermediate scan rates, the PCE difference between the backward (BW) and forward (FW) scans reaches a maximum, as the delayed ionic response can invert the direction of $E_\mathrm{bulk}$, increasing recombination.\cite{cachafeiro2025visualising} At slow scan rates, $E_\mathrm{bulk}$ is screened at every applied voltage because ions have time to respond fully.  The maximum hysteresis and the effect of $E_\mathrm{bulk}$ screening become more pronounced in the devices with thinner C$_{60}$ layer, due to changes in recombination properties ($V_\mathrm{OC}$ statistics in Fig.\,S16).

The TIC signal, which should be largely independent of recombination, depends strongly on $d_{\rm ETL}$ (Fig.\,\ref{fig5}b): both $J_\mathrm{initial}$ and the decay duration decrease with thicker C$_{60}$, consistent with capacitance-limited behavior in which $C_\mathrm{TL,geo}$ scales with $1/d_\mathrm{ETL}$. This rules out a higher $N_\mathrm{ion}$ as the origin of the larger hysteresis and lower slow-scan PCE in the thinner-$\text{C}_{60}$ devices (Fig.\,\ref{fig5}a), since more ions would increase $\sigma_{\rm ion}$, shortening the ion-ToF and shifting hysteresis to higher scan rates, opposite to what is observed in Fig.\,\ref{fig5}c. The timescale trends in Fig.\,\ref{fig5}c, which show faster field screening dynamics for thicker-C$_{60}$ devices, instead reflect a decreasing TL capacitance, with a comparable $\sigma_{\rm ion}$ across all samples, rather than a change in ionic properties.\cite{courtier2019transport} Indeed, the approximated ion-ToF values and peak hysteresis scan rates scale $\propto 1/d_{\mathrm{ETL}}$, similarly to $\tau_\mathrm{RC} = R_{\mathrm{ion}} \cdot C_{\mathrm{TL,geo}}$ (red dashed line in Fig.\,\ref{fig5}c), using $\varepsilon_{\mathrm{ETL}} = 5$ and the $\sigma_{\mathrm{ion}}$ shown in Fig.\,\ref{fig5}d, obtained from extrapolating the trend in $J_\mathrm{initial}$ as a function of $d_{\rm ETL}$. 

Integrating the TIC via Eq.\,\eqref{eq:Nion} with $f_d=1$ yields a $N_{\rm ion}$ that decreases monotonically with $\text{C}_{60}$ thickness from 2.5 to 25\,nm (statistical data in Fig.\,\ref{fig5}e). For comparison, simulations now employ an asymmetric device model to generalize the findings and better represent the p–i–n architecture (see Table\,S2 in SI for parameter details). Similar to the generic simulations discussed above, varying C$_{60}$ thickness ($d_{\rm ETL}$) without considering a geometric $f_d$ correction influences the computed $N_{\rm ion}$ in a way that depends on whether the operating regime is ion- or capacitance-limited (Fig.\,\ref{fig5}f). In the former case, $N_{\rm ion}$ should be independent of $d_{\rm ETL}$, since the integrated charge reflects the full mobile ion population and $f_d$ is unchanged, whereas the latter predicts a dependence of $N_{\rm ion}$ on $d_{\rm ETL}$ due to the electrostatic effect of the C$_{60}$ ETL.

The capacitance-limited $N_\mathrm{ion,max}$, evaluated from Eq.\,\eqref{eq: Nion_max} by assuming the full $V_\mathrm{pre}$ drops across the C$_{60}$ ETL (neglecting the SAM), is shown as the red line in Fig.\,\ref{fig5}e\&f. The experimental $N_\mathrm{ion}$ lies slightly below this limit, but qualitatively follows the trend of $N_\mathrm{ion,max}$. 

We note that the regime cannot be judged from the absolute distance to this line alone, since $N_\mathrm{ion,max}$ is only an upper estimate (see also Fig.\,\ref{fig5}f, where ion-limited curves also lie below it) and a possible decrease due to the opposing displacement current has not been taken into account. The distinguishing signature is instead the dependence on $d_\mathrm{ETL}$: in the ion-limited regime, the computed $N_\mathrm{ion}$ would be independent of the C$_{60}$ thickness; whereas the shown dependency of the measured $N_\mathrm{ion}$ (following the trend of $N_\mathrm{ion,max}$), confirms the TIC is predominantly limited by the ETL capacitance.

Thus, to quantify ionic properties of capacitance-limited devices, we applied the same trend fitting and extrapolation procedure used for the simulated thickness sweep in Fig.\,\ref{fig4}, to both the measured initial current ($J_\mathrm{initial}$) and the average quantified $N_{\mathrm{ion}}$ as a function of $d_{\mathrm{ETL}}$ (black dashed curves in Fig.\,\ref{fig5}d\&e). Because the hole-selective contact in this architecture is only a thin self-assembled monolayer (SAM), across which the potential drop is assumed negligible, the values obtained by extrapolating the fits to $d_{\mathrm{ETL}}=0$, can be taken as good approximations of the TL-free device response. The BCP buffer layer evaporated on top of C$_{60}$ ($\sim$5\,nm, below SEM resolution) could introduce some uncertainty or slight underestimation, since its thickness is comparable to that of the thinnest C$_{60}$ layers. The SAM and BCP layers were used across all devices to preserve a similar $V_{\rm bi}$ throughout the series.\cite{angus2024understanding} Nevertheless, BCP-free devices were confirmed to yield comparable results, as detailed below. We reiterate that at high input $N_\mathrm{ion}$ the deviation below the capacitive limit may become too small for a reliable extrapolation to the true TL-free value (Fig.\,\ref{fig5}f); in practice, C$_{60}$ thicknesses as low as 1\,nm have been demonstrated,\cite{liu2018impact} which would further improve the approach.

The ionic conductivity is obtained from Eq.\,\eqref{eq:Jion0} as $\Sigma \sigma_\mathrm{ion} = J_\mathrm{initial}$(extrapolated to $d_\mathrm{ETL}=0$) $\cdot (V_\mathrm{pre}/d_\mathrm{PSK})^{-1}$, where $V_\mathrm{pre} = 1$\,V (assumed close to $V_\mathrm{bi}$) and $d_\mathrm{PSK} = 600$\,nm. As summarized in Table\,\ref{table: fitted results of pin}, the extrapolation yields $\Sigma \sigma_\mathrm{ion} \approx 7 \times 10^{-9}$\,S\,cm$^{-1}$, in line with previous reports for lead iodide perovskites, \cite{yang2015significance} though reported values span a wide range,\cite{yantara2024toolsets} possibly reflecting the bottlenecks imposed by the different TLs used across the literature.

The same extrapolation for the integrated $N_{\rm ion}$ in Fig.\,\ref{fig5}e yields $3.5\times10^{17}$ cm$^{-3}$, which corresponds to the two-ion case assuming $f_d \to 1$. Together these give an effective ion mobility of $1.3\times10^{-7}$ cm$^2$ V$^{-1}$ s$^{-1}$, close to a previous estimate for comparable devices.\cite{bertoluzzi2018situ} In the SI, we apply the same fitting procedure to an additional batch of p-i-n devices and to a separate set using a different perovskite absorber composition (Fig.\,S17-S19 and Table\,S5). 
For the latter p–i–n devices, an additional batch was fabricated, in which the perovskite absorber was directly contacted by Cu (ITO / 4PADCB (SAM) / Cs$_{0.2}$\allowbreak FA$_{0.8}$\allowbreak Pb(I$_{0.6}$\allowbreak Br$_{0.4}$)$_{3}$ (1.77\,eV) / Cu), effectively minimizing the electrostatic constraint imposed by the ETL and BCP buffer layer. The $N_{\mathrm{ion}}$ obtained for the ETL-free devices is in close agreement to the values obtained from extrapolating the thickness trends, thereby supporting the validity of our method (Fig.\,S20).

If the one-ion scenario is instead assumed, the extrapolated $N_\mathrm{ion}$ represents a lower bound that requires an additional correction $N_\mathrm{ion}/f_d$, with $f_d < 1$. Following the same procedure as in the simulated one-ion TL-free model, the depletion width was approximated ($w_\mathrm{dep}$, Eq.\,\eqref{eq: W_dep_PSK}) assuming $\epsilon_\mathrm{ETL}=5$, $\epsilon_\mathrm{PSK}=44$ and $V_\mathrm{bi}=V_\mathrm{pre}$, where $\epsilon_{\mathrm{PSK}}$ was extracted from the frequency-dependent capacitance of a reference p-i-n sample (Fig.\,S21). This modifies the ion density and ion mobility to $1.8\times10^{18}\,\textrm{cm}^{-3}$ and $2.4 \times 10^{-8}$\,cm$^2$\,V$^{-1}$\,s$^{-1}$, respectively, as listed in Table\,\ref{table: fitted results of pin} for the one-ion model results.

The overall ionic conductivity $\Sigma \sigma_\mathrm{ion}$, derived from the extrapolated $J_\mathrm{initial}$ alone, remains unaffected by the choice of ionic model. This means the true $\mu_\mathrm{ion}$ and $N_\mathrm{ion}$ could be obtained by performing an additional measurement to extract either, like impedance spectroscopy\cite{elhorst2025determining} or low-frequency Mott-Schottky\cite{diethelm2025probing,diekmann2023determination} to extract $\mu_\mathrm{ion}$ and $N_\mathrm{ion}$, respectively. However, these measurements can be limited by the TLs in the same way as TIC, requiring again a similar approach to account for the effect of the TLs.\cite{schmidt2025characterization}

\begin{table}[t!]
    \centering
    \renewcommand\arraystretch{1.4}
    \caption{Experimental results for p-i-n devices under the two- and one-ion models. In the two-ion case $\mu_\mathrm{ion}$ represents the sum of the cation and anion mobilities, i.e. $\mu_\mathrm{ion} = \Sigma\sigma_\mathrm{ion}/(q\,N_\mathrm{ion}) = \mu_\mathrm{c}+\mu_\mathrm{a}$ (total conductivity divided by per-species density, equal concentrations of cations and anions assumed), assuming both species can reach their equilibrium distributions during the transients. $^*$In the one-ion case the TL-free $N_\mathrm{ion}$ is corrected by $f_d = w_\mathrm{dep}/d_\mathrm{PSK}$ (Eq.\,\eqref{eq: W_dep_PSK}).}
    \resizebox{\textwidth}{!}{%
    \begin{tabular}{cccc}
    \toprule
        & Extrapolated $\Sigma \sigma_\mathrm{ion}$ (S cm$^{-1}$) & Extrapolated TL-free $N_\mathrm{ion}$ (cm$^{-3}$) & Computed ion mobility $\mu_\mathrm{ion}$ (cm$^2$ V$^{-1}$ s$^{-1}$) \\
    \midrule
        \textbf{Two-ion}& $7\times10^{-9}$ & $3.5\times10^{17}$  &  $1.3\times10^{-7}$ \\
        \textbf{One-ion} & $7\times10^{-9}$ & $1.8\times10^{18}\,^{*}$ &  $2.4 \times 10^{-8}$ \\
    \bottomrule
    \end{tabular}%
    }
    \label{table: fitted results of pin}
\end{table}

\subsubsection{HTL doping variation of n-i-p devices}

Having demonstrated with the p-i-n devices the effect of varying the TL geometric capacitance through $d_{\mathrm{ETL}}$, we now turn to the effect of TL doping through $N_{\mathrm{HTL}}$, which is expected to modulate $C_\mathrm{TL}$ through the depletion capacitance (Eq.\,\eqref{eq: C_dep}). The same perovskite composition is used in the n-i-p devices: FTO/SnO$_2$/Cs$_{0.05}$\allowbreak (FA$_{0.95}$\allowbreak MA$_{0.05}$)$_{0.95}$\allowbreak Pb(I$_{0.95}$\allowbreak Br$_{0.05}$)$_3$ (1.56\,eV)/Spiro-OMeTAD: Li-TFSI/Au, where the Li-TFSI doping concentration is tuned in Spiro-OMeTAD, which is a standard route to adjust its hole conductivity.\cite{kong2021co2} Cross-sectional SEM confirms a perovskite absorber thickness ($d_\mathrm{PSK}$) of approximately 700\,nm, with Spiro-OMeTAD HTL and SnO$_2$ ETL thickness of approximately 150\,nm and 10\,nm, respectively (Fig.\,S
22). 

\begin{figure}[t]
\centering
  \includegraphics[width=\linewidth]{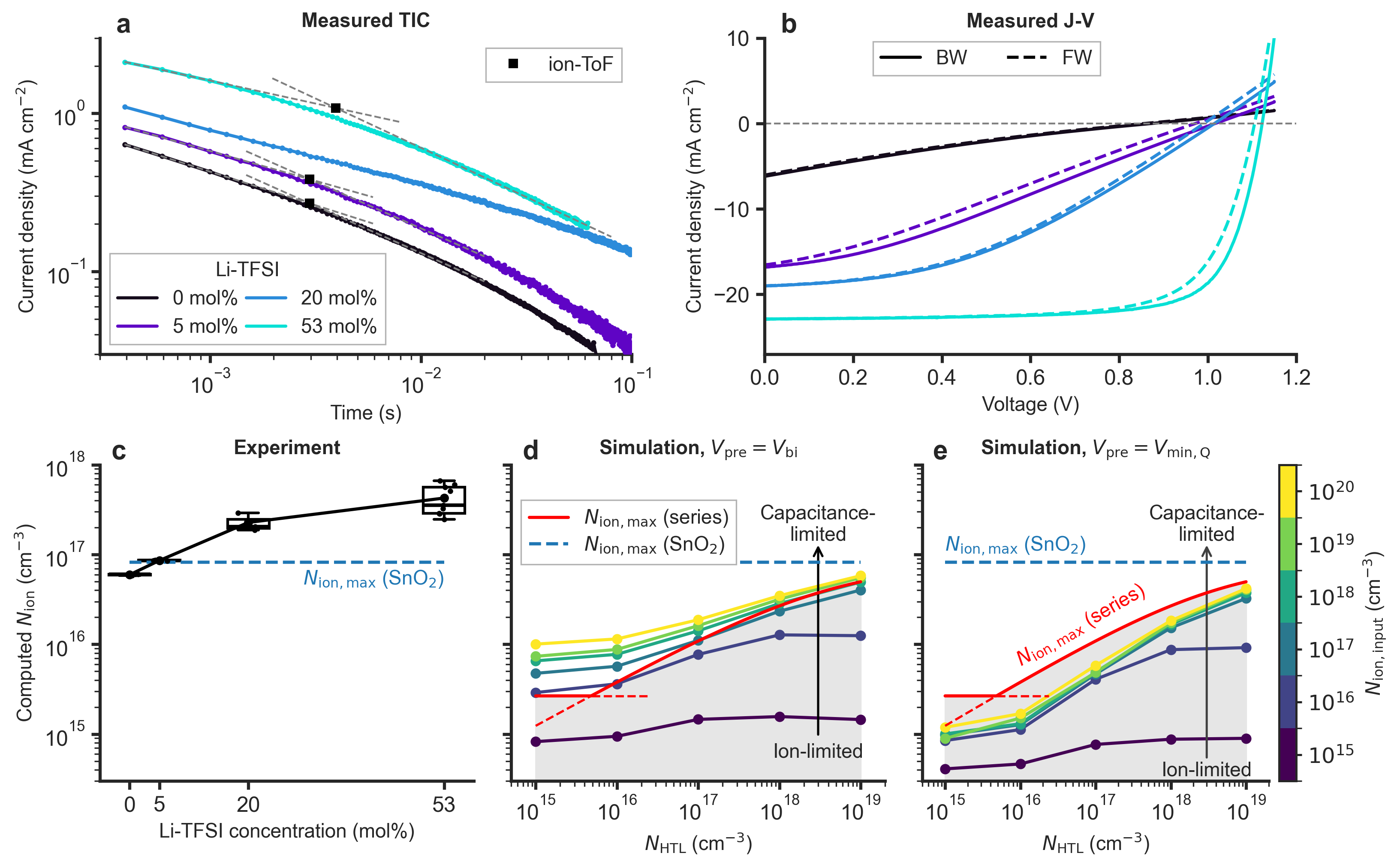}
  \caption{Experimental validation and simulation of transport-layer-dependent ionic quantification for n-i-p samples. (a) Measured TIC, (b) $J$-$V$ curves for n-i-p architecture ($\text{Au}/\text{Spiro-OMeTAD}/\text{Perovskite}/\text{SnO}_{2}/\text{FTO}$) with varied Li-TFSI dopant concentration in Spiro-OMeTAD (HTL), and (c) $N_{\mathrm{ion}}$ computed from TIC with $f_d=1$. Simulated results for the n-i-p model across various Spiro-OMeTAD p-doping concentrations ($N_{\mathrm{HTL}}$) and input ion densities, preconditioned at (d) $V_{\mathrm{pre}} = V_{\mathrm{bi}}$
  and (e) $V_{\mathrm{pre}} = V_{\mathrm{min},Q}$. 
  The red solid line is the series combination of the Spiro-OMeTAD (HTL) and SnO$_2$ (ETL) capacitive limits (Eq.\,\ref{eq:Nion_max_asym}); the blue dashed line shows the SnO$_2$-only limit. The red dashed lines highlight the Spiro-OMeTAD geometric ($C_\mathrm{TL,geo}$) and $N_{\mathrm{HTL}}$-dependent effective ($C_\mathrm{TL,eff}$, Eq.\,\eqref{eq: C_dep}). The simulations utilize the parameters detailed in Table\,S3.}
  \label{fig6}
\end{figure}

The measured TIC and the 100 mV\,s$^{-1}$ $J$–$V$ performance under $\approx$ 1 sun intensity (AM1.5G solar simulator) are shown in Fig.\,\ref{fig6}a\&b, both showing a significant dependence on the Li-TFSI doping concentration $N_{\rm HTL}$ due to its modulation of the HTL conductivity. The $N_{\rm ion}$ quantified from TIC measurements via Eq.\,\eqref{eq:Nion} ($f_d=1$) rises with Li-TFSI concentration (Fig.\,\ref{fig6}c), reproducing qualitatively also the $N_{\mathrm{TL}}$-dependent trends predicted by the simulations above.

Notably, the quantified $N_\mathrm{ion}$ is overall significantly higher than for the p-i-n samples, already reaching $> 10^{17}$\,cm$^{-3}$ without any corrections, which is commonly above the typical detection limits reported in simulations with TLs.\cite{diekmann2023determination,schmidt2025many} This is despite the thick HTL used in the n-i-p configuration ($\sim$150\,nm Spiro-OMeTAD), which should in case of no doping reduce $C_\mathrm{TL}$, $\Delta\phi_\mathrm{PSK}$ and thus the measured $N_\mathrm{ion}$.

Simulations of the corresponding n-i-p model device (parameters in Table\,S3), preconditioned at $V_\mathrm{pre} = V_\mathrm{bi}$ and at $V_{\mathrm{\min},Q}$, are shown in Fig.\,\ref{fig6}d\&e across various Spiro-OMeTAD doping densities ($N_\mathrm{HTL}$) and input ion densities. As expected, a more conductive HTL sustains a narrower depletion width, thus $C_\mathrm{HTL}$ rises with $N_\mathrm{HTL}$ (Eq.\,\ref{eq: C_dep}), increasing the total capacitance and the computed $N_\mathrm{ion}$. 

The capacitive limit $N_\mathrm{ion,max}$ is computed from Eq.\,\ref{eq:Nion_max_asym} using the series combination of asymmetric HTL and ETL capacitances (red lines in Fig.\,\ref{fig6}d\&e), where the HTL depletion capacitance (Eq.\,\ref{eq: C_dep}) is approximated with $\Delta\phi_{\mathrm{HTL}} = V_{\mathrm{bi}}/2$ (the exact series partition depends on $C_{\mathrm{ETL}}/C_{\mathrm{HTL}}$ but has only a small effect on the log scale). Because the SnO$_2$ ETL is much thinner than the Spiro-OMeTAD HTL ($C_\mathrm{ETL} \gg C_\mathrm{HTL}$ at low $N_\mathrm{HTL}$), the series capacitance is dominated by the HTL. In the limit $C_\mathrm{HTL} \to \infty$ (i.e.\ very high doping), the series expression tends to become limited by $C_\mathrm{ETL}$ instead, shown as the blue dashed lines in Fig.\,\ref{fig6}c-e. With preconditioning at $V_\mathrm{bi}$ (Fig.\,\ref{fig6}d), the computed $N_\mathrm{ion}$ can surpass the simple capacitance-limit due to the contribution of the initial discharge process resulting from ionic accumulation of reversed polarity above $V_{\mathrm{\min},Q}$, as discussed before. Nevertheless, the computed $N_\mathrm{ion}$ remains closer to the series capacitance limit when this effect is prevented, i.e. by setting $V_\mathrm{pre} = V_{\mathrm{\min},Q}$ in the simulations (Fig.\,\ref{fig6}e).
  
Although the distinction between the ion- and capacitance-limited regimes is in this case somewhat less obvious, the simulated $N_\mathrm{HTL}$ dependence qualitatively reproduces the experimental trend, confirming that TL doping also modifies the $N_{\mathrm{ion}}$ determined from TIC. In Fig.\,\ref{fig6}a, the ion-ToF is difficult to define and is only roughly approximated here, again from the intersection of straight-line fits to the early- and late-time segments of the TIC.\cite{peterson2025reliable} Under the assumption that at the highest Li-TFSI concentration (53~mol\%) the HTL is heavily doped so that the series capacitance is mainly limited by the SnO$_2$ ($C_{\mathrm{TL}} \approx C_{\mathrm{SnO_2}} \approx 8\times10^{-7}$~F~cm$^{-2}$), the approximate $\tau \approx 4$~ms gives $\Sigma \sigma_{\mathrm{ion}} = d_{\mathrm{PSK}} \, C_{\mathrm{SnO_2}} / \tau \approx 1.4\times10^{-8}$~S~cm$^{-1}$, which is somewhat higher but comparable to the p-i-n value in Table\,\ref{table: fitted results of pin}. Nevertheless, for all but the undoped samples, the measured $N_\mathrm{ion}$ even exceeds the capacitive limit estimated for the SnO$_2$ ETL (blue dashed line in Fig.\,\ref{fig6}c). The computed ionic charge being above the theoretical detection threshold indicates additional effects may be at play, which we consider next. 

While mobile ionic charges are commonly assumed to remain confined within the perovskite absorber, extensive experimental evidence demonstrates ionic interdiffusion across the PSC layers. This is particularly the case with organic HTLs such as Spiro-OMeTAD.\cite{kerner2021organic,besleaga2016iodine,shen2024ultrathin} Previous reports have discussed iodine invasion into Spiro-OMeTAD,\cite{wang2022transporting} Li$^+$ penetration into the perovskite and ETL,\cite{li2017extrinsic,xu2024ion,choi2023irreversible} and metallic ion diffusion.\cite{li2017extrinsic,domanski2016not,hu2025electrolytic} These effects are generally linked to degradation and stability issues in longer timescales, with the associated PCE losses sometimes being partially reversible.\cite{domanski2017migration} This points to broader considerations beyond the confined-ion model: one in which mobile ionic charges reversibly redistribute in and out of the TLs, thereby overcoming the limitation of TLs, or one in which the TIC depends on additional electrochemical processes.\cite{xu2024beyond}

To test experimentally whether ionic species can move reversibly into the TLs under bias, on timescales relevant to typical TIC measurements, we performed time-of-flight secondary-ion mass spectrometry (ToF-SIMS) with in-situ biasing\cite{harvey2025applying} on the n–i–p stack. In a fresh device (details in Fig.\,S23), the anion profiles reveal I$^-$ within the Spiro-OMeTAD HTL, with its intensity near the Au electrode continuously increasing over two 1.2--0\,V biasing cycles, indicating bias-induced changes in the distribution of ionic species within the HTL. In an aged device (details in Fig.\,S24), the cation profiles show Li$^+$ at the SnO$_2$ ETL, in line with previous reports\cite{li2017extrinsic}. While our ToF-SIMS measurements cannot establish whether mobile ionic charges migrate reversibly into the TLs, the presence of ionic species across other layers suggests that, in this case, ionic conductivity is perhaps not strictly confined to the perovskite absorber. Alternatively, it may reflect more complex interfacial electrochemistry or space charge effects beyond the simple ionic accumulation at the PSK/TL interfaces assumed in the confined-ion model.\cite{kerner2021organic,xu2024beyond}

\subsection{Above the capacitive limit: model extensions and possible mechanisms}\label{sec:above-limit}
\subsubsection{Ion-permeable charge transport layers}

The confined-ion model above captures a key feature of TIC: the $N_{\rm ion}$ determined from TIC is set mainly by the electrostatic properties of the TLs, rather than by the absorber's intrinsic ion density alone. This model generally reproduces the observed dependence on TL properties. However, the measured $N_{\rm ion}$ in Fig.\,\ref{fig6} for the n-i-p devices (with thick TLs) can exceed not only the TL capacitance limits but even the TL-free extrapolation of the p-i-n case. Additionally, the measured TIC for n-i-p shows a particularly longer decay with a less obvious ion-ToF (Fig.\,\ref{fig6}a), compared to the confined-ion simulations and experimental p-i-n results. To provide plausible explanations for this, we consider two simple extensions of the baseline drift-diffusion model: a permeable-TL model, in which ions are no longer strictly confined to the absorber, and a trapped-state model, in which shallow interfacial defects can capture and release electronic charges, modifying the device electrostatics.
 
\begin{figure}[t!]
\centering
  \includegraphics[width=\linewidth]{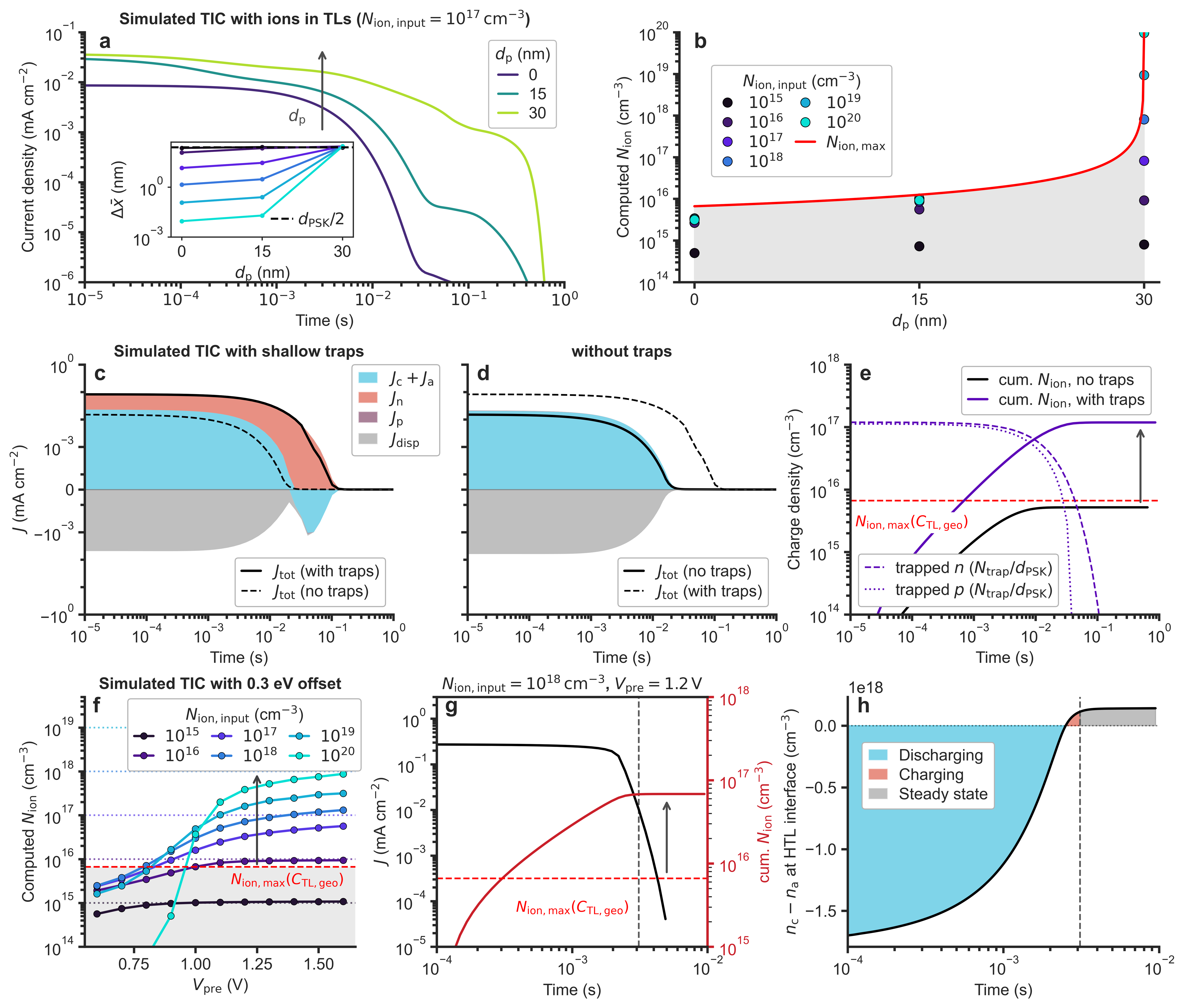}
  \caption{(a) Simulated TIC for different ion penetration depths ($d_p$) into the 30 nm thick TLs under an input ion density of 10$^{17}$ cm$^{-3}$; the inset shows the average ionic displacement ($\Delta\bar{x}$ from Eq.\,\eqref{eq:CoM}) as a function of $d_p$ for various input $N_{\rm ion}$. (b) Computed $N_{\rm ion}$ as a function of $d_p$ for various input $N_{\rm ion}$. Permeable-model simulations are performed using identical TL electronic parameters as in the confined case, with the ionic mobility in the TLs set equal to that of the perovskite absorber. 
  The red curve shows the capacitive limit $N_{\mathrm{ion, max}}$ from Eq.\,\eqref{eq: Nion_max} with the effective geometric capacitance $C_\mathrm{TL,geo}=\epsilon_0 \epsilon_\mathrm{TL}/(d_\mathrm{TL}-d_p)$.
  (c) Simulated TIC with slowly-releasing shallow traps (150\,meV, $10^{14}$\,cm\textsuperscript{-2}) at the HTL/PSK interface (Table\,S4), compared to (d) without traps. (e) Cumulative charge density integral from the transients in (c) and (d).}
  \label{fig7}
\end{figure}

In the permeable model, mobile ionic charges are allowed to enter the TLs, considering a penetration depth $d_p$ into each TL for simplicity, with an ionic mobility equal to that of the perovskite absorber, for both ions. As illustrated in Fig.\,\ref{fig7}a, increasing $d_p$ systematically increases the transient current signal, compared with the confined case ($d_p=0$). Due to the slower process, the ion-ToF becomes more difficult to identify, as for the measured TIC of the n-i-p devices (Fig.\,\ref{fig6}a). The timescales of TL penetration depend on the ionic mobility inside the TL regions: when it equals that of the perovskite, if follows immediately after the initial plateau; however, when the ionic mobility is set significantly lower than in the perovskite, it separates into a distinct tail at much slower timescales (Fig.\,S25), resembling the changes at slower timescales (minutes) previously observed in degradation studies.\cite{thiesbrummel2024ion,torre2025ion} As shown in Fig.\,S25c, the faster part of TIC corresponds to ionic redistribution within the absorber, identical to the confined model, whereas the subsequent stage reflects the migration of the accumulated ions across the interfaces and through the TLs. Multi-step or prolonged TIC decays observed experimentally could therefore be interpreted as signatures of slower processes such as ionic TL penetration.

The role of $d_p$ can be interpreted in terms of the ion- and capacitance-limited regimes discussed above, since it is somewhat analogous to the $d_{\mathrm{TL}}$ sweeps. As ions penetrate the TLs, the inaccessible TL region that the ionic charge cannot reach shrinks to a width of $d_\mathrm{TL}-d_p$, raising the effective geometric capacitance $C_\mathrm{TL,geo}=\varepsilon_0 \varepsilon_\mathrm{TL}/(d_\mathrm{TL}-d_p)$ and thus the effective capacitive limit (red curve in Fig.\,\ref{fig7}b).
As seen in the inset of Fig.\,\ref{fig7}a, increasing $d_p$ therefore raises average ionic displacement $\Delta \bar{x}$ and the computed $N_{\mathrm{ion}}$ (Eq.\,\eqref{eq:Nion}, using $f_d=1$).
Once $d_p=d_{\mathrm{TL}}$ (each TL is 30\,nm here), the entire stack is accessible to the ions and the computed $N_{\mathrm{ion}}$ converges to the true $N_\mathrm{ion,input}$ (Fig.\,\ref{fig7}b), as in the (ion-limited) TL-free two-ion case. Taken with appropriate caution, ionic permeation of the TLs therefore offers a physically plausible explanation for the high $N_{\mathrm{ion}}$ values often reported experimentally from TIC measurements, since the technique can reach higher sensitivity by minimizing the TL region that cannot be accessed by mobile ions.


\subsubsection{Traps at the perovskite/TL interface}

A second explanation for the unexpectedly high charge densities obtained from TIC ($N_\mathrm{ion}$) might be based on more complex electrochemical processes beyond ion migration as currently included in drift-diffusion simulations.\cite{xu2024beyond} As an example, we consider now the formation and slow release of interfacial charge\cite{weber2018formation} \textit{via} electronic defects. To simulate this, interfacial defect states are added, which can trap and release charges during TIC (parameter set is in Table\,S4).

The influence of traps on the simulated TIC depends naturally on their parameters, including trap type and location, depth (energy level), capture rate, etc. Trapped charges can either contribute directly to the transient current upon slow thermal release or simply act as a fixed interfacial charge sheet that modifies the capacitive limit to ionic accumulation at the TL interfaces. Similar model adjustments (adding surface states at TL interfaces) have been previously introduced in the literature to reproduce the large low-frequency capacitance of PSCs.\cite{jacobs2018two}

As an illustrative case, we consider in Fig.\,\ref{fig7}c–e two shallow traps at the PSK/HTL interface, 150\,meV below the CB (electron trap) and above the VB (hole trap). Using both trap types prevents the formation of a net trapped surface charge at the interface during the precondition, such that the initial ion distribution and subsequent ionic current is comparable to the model without traps. Upon switching to 0\,V, if the emission rate is sufficiently slow, the release of trapped electronic charges takes place in the same timescales as ion migration. As shown in Fig.\,\ref{fig7}c, this brings the total current $J_\mathrm{tot}\equiv J$ significantly above the purely ionic current ($J_\mathrm{c}+J_\mathrm{a}$) and the trap-free case (Fig.\,\ref{fig7}d), due to the electrons released by the n-trap at HTL and drifting across the perovskite towards the ETL ($J_n$). For a sufficiently high density of initially occupied trap states, the computed charge density via Eq.\,\ref{eq:Nion} can therefore exceed the capacitive limit as it becomes limited instead to $N_\mathrm{trap}/d_\mathrm{PSK}$ (Fig.\,\ref{fig7}e), where $N_\mathrm{trap}$ is the surface density of traps filled during the precondition.

While disentangling the role of charge trapping and ion migration on the transient response of PSCs remains relevant,\cite{van2022slow,futscher2021defect,veurman2024deciphering} it is strongly suggested that in practice these processes may be coupled, meaning PSCs should be considered as solid-state electrochemical cells with respect to ion migration effects,\cite{xu2024beyond,hu2025electrolytic} i.e. by including electrochemical terms in drift-diffusion for the generation and annihilation of ionic charge.\cite{bertoluzzi2021incorporating,bitton2023perovskite} Therefore the essential conclusion with regards to the role of electronic traps on TIC is that slow processes releasing significant amounts of electronic charge on the ion migration timescales, for instance due to interfacial reactions\cite{fremouw2026electrochemical} or ions changing their charge state,\cite{tyagi2025tracing} could also inflate the computed $N_{\mathrm{ion}}$ by orders of magnitude above the capacitive limit.

\subsubsection{Discharge of initial ionic accumulation}
This last scenario is not an extension to the baseline model, but a slightly more detailed look at the effect of preconditioning above $V_{\mathrm{min},Q}$ as briefly discussed in Fig.\,\ref{fig2}h. As defined earlier, $V_{\mathrm{min},Q}$ is the applied voltage that minimizes the net ionic charge (per unit area) in the perovskite, with
\begin{equation}
Q_{\mathrm{ion}}(V)
  \;=\; q\!\int_{0}^{d_{\mathrm{PSK}}}\!
        \bigl|c(x)-a(x)\bigr|\,\mathrm{d}x.
\end{equation}

Because the PSK is mostly field-free at this voltage it is also referred to as the PSK flat-band condition ($V_{\mathrm{flat}}$)\cite{hart2024more} or `ion-free' voltage.\cite{torre2025ionic} Above $V_{\mathrm{min},Q}$, the bulk electric field in PSK switches direction, with ionic accumulation at the TL interfaces correspondingly switching polarity. The capacitive limit for $N_{\mathrm{ion}}$ as in Eq.\,\ref{eq: Nion_max}-\ref{eq:Nion_max_asym} assumes the dominant process contributing to TIC is the charging of the TLs as ions accumulate. However, if $V_{\mathrm{pre}}$ is substantially above $V_{\mathrm{min},Q}$ for the TIC measurement, the initial signal may originate instead from the discharge of the pre-accumulated ions, of switched polarity. Under forward bias, if there is a high barrier at the TL/PSK interface to electronic injection into PSK, the electronic carrier density in the TLs will be substantial, allowing for higher ionic accumulation than the capacitive limit derived for the charging process. This results in a computed $N_{\mathrm{ion}}$ above the capacitive limit, as shown in Fig.\,\ref{fig2}h and illustrated further in Fig.\,\ref{fig7}f for the case of highest energetic offset at the TL/PSK interface: $\Delta E_{\rm TL}=0.3\,\mathrm{eV}, V_{\rm bi}=0.8\,\mathrm{V}, V_{\mathrm{min},Q}\approx0.55\,\mathrm{V}$. Depending on $V_{\mathrm{pre}}$ and the input ion density, $N_{\mathrm{ion}}$ from TIC can surpass the TL charging limit by almost two orders of magnitude. A selected TIC trace is shown in Fig.\,\ref{fig7}g for the 1.2\,V precondition and $10^{18}$\,cm\textsuperscript{-3} input $N_{\mathrm{ion}}$. The measured current and thus cumulative charge density integral originates mainly from the discharging process, as shown in Fig.\,\ref{fig7}h. The net ionic charge at the HTL interface is initially negative and decreases over time as the accumulated anions discharge into the bulk. The charging process, starting when the polarity of ionic accumulation at the interface switches sign, quickly reaches a steady-state and only contributes minimally to the charge integral, as the final state accumulation is more strongly limited than the forward bias-preconditioned initial state. Thus, injection barriers under forward bias above $V_{\mathrm{min},Q}$ provide yet another mechanism by which the computed $N_{\mathrm{ion}}$ from TIC in experiment may be substantially higher than the amount of ions needed to charge the TL capacitors and screen the built-in bulk electric field, due to the measurement being dominated instead by the discharge process of initially accumulated ions.

\section{Conclusion}

In summary, this work provides a comprehensive mechanistic evaluation of the voltage step-induced transient ion current (TIC), or BACE, for quantifying ionic parameters in PSCs. Through a combination of drift-diffusion simulations and experimental validation, we demonstrate that the ion density determined from TIC is fundamentally constrained by the electrostatic properties of TLs, and does not necessarily reflect the intrinsic ionic properties of the perovskite absorber but the capacitance of the TLs. Experiments with systematically varied C$_{60}$ ETL thickness (p-i-n) and Spiro-OMeTAD HTL p-doping (n-i-p) confirm the strong dominance of the TL properties, across different PSC architectures.

To overcome this bottleneck, we show that a correction based on the average ionic displacement is the most important factor for accurately quantifying the ion density from TIC. The average ionic displacement decreases below half of the perovskite thickness once the ions can screen the bulk electric field, requiring a correction factor ($f_d$) to adequately calculate $N_{\mathrm{ion}}$ from the charge-density integral (Eq.\,\eqref{eq:Nion}). 
Building on this, we show that extrapolating a TL-thickness series towards the TL-free reference ($C_\mathrm{TL} \to \infty$) provides a practical experimental route to quantify ionic conductivity, density, and mobility of the perovskite absorber. 
Applied to the ETL of our p-i-n samples, the thickness-extrapolation approach yields the true ionic conductivity and allows us to estimate density and mobility. Varying the TL thickness thus provides a route to quantify ionic parameters in the capacitance-limited regime: the ionic conductivity $\sigma_\mathrm{ion}$ is quantified without limitation, since it is independent of the ionic displacement factor; whereas the ion density $N_\mathrm{ion}$ is obtained only as a lower estimate, since within the single-mobile-ion model a residual $f_d$ correction may still be required at high ion densities, even in the TL-free situation. For this case, we propose a calculation of $f_d$ based on the equilibrium ionic depletion width in perovskite, which recovers the true ion density.

To address the anomalously high ion densities frequently reported in the literature and observed in our thick-TL n-i-p samples, we simulated different mechanisms by which the capacitive limit can be surpassed. In the permeable model, ions are no longer confined to the absorber but penetrate the TL stack, raising the capacitive limit by shrinking the TL region inaccessible by the ions. In the trapped-state model, the slow release of shallow interfacial defects provides a source of electronic current during TIC that inflates the computed ion density beyond the ion drift current signal alone. Additionally, forward bias preconditioning, above the voltage at which the bulk electric field switches direction, can result in the measured current originating from the discharge of initially accumulated ions under a higher capacitive limit influenced by external charge injection and the height of the injection barrier at the TL/PSK interface. All of these mechanisms can raise the upper limit on the inferable $N_\mathrm{ion}$ from TIC.

Consequently, changes in TIC do not necessarily reflect a genuine increase in the mobile ion density: unless accompanying changes in TL properties are accounted for, they may instead reflect only the electrostatics of TLs, especially in the capacitance-limited regime, where the quantified $N_\mathrm{ion}$ is set by the charge the TL capacitance can accommodate at the interfaces rather than by the absorber's intrinsic ion density. The true ion density can be cleanly quantified from TIC only when the mean displacement of mobile ions in the perovskite bulk (the $f_d$ term) can be reliably approximated.

Our findings provide essential guidelines for accurately quantifying ionic properties from TIC measurements, a critical step toward reliable ion-parameter extraction through electrical measurements in perovskite optoelectronics.

\section{Methodology}
\subsection{Drift-diffusion simulations}
Device simulations were performed using the commercial drift-diffusion solver Setfos (Fluxim AG). A full list of material parameters and boundary conditions is provided in the Supporting Information (Tables.\,S1-S4). In all baseline simulations, both anionic and cationic species were included with identical ion densities and mobilities to isolate the electrostatic effects of ionic redistribution from asymmetric transport effects, unless where the one-ion model is explicitly specified. To elucidate the fundamental role of TLs, we first employed a confined symmetric device model, in which the ETL and HTL possess identical material properties (for electrons and holes, respectively) and the mobile ions are confined within the perovskite absorber. This configuration eliminates structural asymmetry and allows a simpler evaluation of how TL thickness ($d_\mathrm{TL}$), dielectric constant ($\varepsilon_\mathrm{TL}$), doping density ($N_\mathrm{TL}$), and band alignment modulate the capacitive bottleneck in the transient ion current (TIC). All $d_\mathrm{TL} \to 0$ extrapolations of the simulated data were performed using monotonic cubic splines fitted to the sweep points and evaluated at $d_\mathrm{TL}=0$.

To generalize the trends, the symmetric model was adapted into asymmetric p-i-n and n-i-p device models more closely reproducing the fabricated architectures, with the corresponding parameters summarized in Table\,S2\&S3. In the permeable model, mobile ions are allowed to penetrate a depth $d_p$ across the perovskite/TL interfaces with an ionic mobility equal or lower to that of the absorber, while keeping all electronic and TL parameters identical to the confined baseline. In the trapped-state model, shallow electronic trap states were introduced at the perovskite/HTL interface (Table\,S4). 

\subsection{Device fabrication}
Two device architectures were fabricated to experimentally validate the influence of TL properties on the quantified ion density.

\textit{p-i-n architecture} (ITO/4PADCB/perovskite/C$_{60}$/BCP/Cu):
Patterned ITO-coated glass substrates were sequentially cleaned by ultrasonication in detergent, deionized water, acetone, and isopropanol, dried under N$_2$ flow, and treated with UV–ozone. The hole-transport layer was formed by spin-coating a 0.3\,mg\,mL$^{-1}$ 4PADCB solution in ethanol, preheated at $40\,^{\circ}\text{C}$, onto the cleaned ITO substrates after 30\,s of resting on the substrate. 
The substrates were then spun at 3000\,rpm for 30\,s and annealed at $100\,^{\circ}\text{C}$ for 2\,min. The 1.56\,eV perovskite precursor solution was prepared at a concentration of 1.5\,M in DMF/DMSO (4:1, v/v), with a nominal stoichiometry of Cs$_{0.05}$\allowbreak (FA$_{0.95}$\allowbreak MA$_{0.05}$)$_{0.95}$\allowbreak Pb(I$_{0.95}$\allowbreak Br$_{0.05}$)$_3$, containing 5\,mol\% excess PbI$_2$ and 25\,mol\% MACl. After overnight stirring, 45\,uL of the precursor solution was spin-coated using a two-step program of 2500\,rpm for 10\,s and 5000\,rpm for 40\,s, with 300\,uL chlorobenzene antisolvent dropped 30\,s before the end of the second step. The films were then annealed at $100\,^{\circ}\text{C}$ for 30\,min. 
C$_{60}$ was thermally evaporated with nominal thicknesses of 2.5, 5, 10, and 25\,nm, followed by deposition of a thin BCP buffer layer. The devices were completed by thermally evaporating 150\,nm Cu electrodes through a shadow mask.
For the control devices of I, the perovskite absorber was prepared using a 1.2\,M Cs$_{0.2}$\allowbreak FA$_{0.8}$\allowbreak Pb(I$_{0.6}$\allowbreak Br$_{0.4}$)$_{3}$ precursor solution in DMF/DMSO (4:1, v/v). The precursor solution was filtered through a 0.22\,um PTFE filter and spin-coated onto the ITO/4PADCB substrates using a two-step program of 2000\,rpm for 10\,s and 6000\,rpm for 40\,s, with 300\,uL chlorobenzene dropped 20\,s after the start of the second step. The films were annealed at $60\,^{\circ}\text{C}$ for 2\,min and then at $100\,^{\circ}\text{C}$ for 7\,min. The same C$_{60}$/BCP/Cu deposition sequence was then used to complete the 1.77\,eV bandgap devices.

\textit{n-i-p architecture} (FTO/SnO$_{2}$/perovskite/Spiro-OMeTAD:Li-TFSI/Au):
FTO substrates were cleaned as described for the ITO substrates. A diluted SnO$_2$ dispersion (15\,wt\%, Sigma-Aldrich; Sn$O_2$:H$_2$O = 1:3\,v/v) was filtered (0.45\,um filter), spin-coated at 3000\,rpm for 30\,s, and annealed at $180\,^{\circ}\text{C}$ for 20\,min in air. The perovskite precursor (1.5 M, Cs$_{0.05}$\allowbreak (FA$_{0.95}$\allowbreak MA$_{0.05}$)$_{0.95}$\allowbreak Pb(I$_{0.95}$\allowbreak Br$_{0.05}$)$_3$ with 5\,mol\% excess PbI$_2$ and 25\,mol\% MACl) was prepared in DMF/DMSO (4:1\,v/v). A 45\,uL of the precursor solution was spin-coated using a two-step program (2500\,rpm for 10\,s, then 5000\,rpm for 40\,s), with chlorobenzene dropped 30\,s before the end of the second step, followed by annealing at $100\,^{\circ}\text{C}$ for 30\,min. Spiro-OMeTAD solutions were prepared by dissolving 57.8\,mg of Spiro-OMeTAD in 800\,uL of chlorobenzene. Different doping concentrations were obtained by adding 0, 3.7, 14.8, or 37\,uL of a mixed dopant solution composed of 1.645\,mL of 4-tert-butylpyridine (tBP) and 1\,mL of Li-TFSI solution (520\,mg\,mL$^{-1}$ in acetonitrile), corresponding to Li-TFSI doping levels of 0, 5, 20, and 53\,mol\% relative to Spiro-OMeTAD, respectively. Subsequently, 35\,uL of the corresponding Spiro-OMeTAD solution was spin-coated onto the perovskite films at 3000\,rpm for 30\,s. Finally, 60\,nm Au electrodes were thermally evaporated, and the devices were stored in dry air for at least 36 h before characterization.

\subsection{Measurements}
\subsubsection*{Cross-sectional scanning electron microscopy (SEM)}

Focused ion beam (FIB) cross-sections were used to verify the 1.56\,eV perovskite thicknesses, whereby $d_{\mathrm{PSK}} \approx 600$\,nm for p-i-n (Fig.\,S15) and $\approx 700$\,nm for n-i-p (Fig.\,S22), as well as the systematic variation of the C$_{60}$ layer in the p-i-n series (Fig.\,S15). The FIB cross-sections were prepared on a TFS Helios 5 CX machine. First, a $\leq$100\,nm layer of carbon was deposited on the electrode surface using a 2\,kV\,@\,0.55\,nA\,um$^{-2}$ electron beam to protect thin electrodes from ion beam damage. Then a 500\,nm tungsten layer was deposited using a 30\,kV\,@\,8\,pA\,um$^{-2}$ gallium ion beam to reduce curtaining artefacts. The ion beam milling was performed using a 30\,kV\,@\,80\,pA regular cross-section, followed by a 30\,kV\,@\,80\,pA cleaning cross-section. The SEM images were acquired using the TLD-SE detector in immersion mode with 2\,kV\,@\,86\,pA electron beam settings and with 4$\times$ line integration, resulting in a total effective dwell time of 2\,us per pixel to reduce sample charging and beam damage during imaging. The images were then post-processed using non-local means denoising\cite{roels2020interactive} and contrast enhancement. A cleaved cross-sectional SEM image (Fig.\,S18) was also acquired to verify the 1.77\,eV perovskite thickness for the p-i-n architecture, found to be $\approx 500$\,nm.  The secondary electron SEM image was acquired at 5\,kV.

\subsubsection*{Transient ion current (TIC) Measurements}
TIC measurements (also known as BACE for bias-assisted charge extraction) were performed in the dark at room temperature using a BioLogic SP-300 potentiostat (BioLogic Science Instruments). Devices were first preconditioned at a constant voltage $V_\mathrm{pre}$ (expected to be close to the built-in voltage, $V_\mathrm{pre}\approx1\,\mathrm{V}$) for 300\,s, to ensure the ionic distribution reached steady state, after which the bias was abruptly switched to 0 V and the resulting transient current was recorded. The initial fast electronic current spike was excluded from the analysis, and the contribution of the displacement current was assumed to be small in the TL-free limit. The ionic charge $Q$ was obtained by integrating the transient current density. The mobile ion density was calculated using Eq.\,\eqref{eq:Nion}, and intrinsic ionic parameters were obtained by fitting the $d_\mathrm{TL}$-dependent trends with monotonic cubic splines and extrapolating to the TL-free limit ($d_\mathrm{TL}=0$).

\subsubsection*{Current-voltage and hysteresis characterization}
Current density-voltage ($J$-$V$) curves were measured under $\approx$\,1~sun illumination (AM1.5G solar simulator) in both forward and reverse scan directions, using the BioLogic SP-300 potentiostat with a class ABB AM1.5G solar simulator (LCS-100, 94011A, ORIEL, USA). Scan-rate-dependent measurements were performed to compare the scan rate of maximum $J$-$V$ hysteresis with the ion time-of-flight obtained from TIC.

\subsubsection*{Time-of-flight secondary-ion mass spectrometry (ToF-SIMS)}
ToF-SIMS measurements were performed using a ToF-SIMS V instrument (IONTOF). A 25\,keV Bi$_{3}^{+}$ primary ion beam was used for analysis over a $50 \times 50$\,$\mu$m$^2$ raster area. Depth profiling was conducted using a 1\,keV Cs$^{+}$ sputtering beam over a $150 \times 150$\,um$^2$ raster area. The collected secondary-ion intensities were normalized to the total ion counts. For biased ToF-SIMS measurements, the sample was mounted on a biasing holder. A preset bias voltage was applied for 10\,min before depth profiling and was maintained throughout the ToF-SIMS acquisition. A complete depth profile required approximately 30\,min, during which the applied bias was continuously held.

\medskip
\textbf{Supporting Information} \par 
Supporting Information (SI) is available online or from the author. A list of abbreviations, symbols and variables used throughout this work is available in Table S6 in the SI file.

\medskip
\textbf{Acknowledgements} \par 
This research received funding from the European Union's Horizon 2020 program under grant agreement no. 851676 (ERC StGrt) and SNF Korean-Swiss Science and Technology Programme project number 218518. The authors gratefully acknowledge ScopeM for their support \& assistance in this work.

\medskip

\bibliographystyle{MSP}
\bibliography{reference}

\end{document}